\documentclass{article}
\pdfoutput=1

\usepackage[utf8]{inputenc}
\usepackage{amsmath,amsthm}
\usepackage{amsfonts}
\usepackage{amssymb}
\usepackage{epic,eepic,latexsym,color}
\usepackage{ifthen,graphics,epsfig}
\usepackage[section]{algorithm}
\usepackage[noend]{algorithmic}

\usepackage{xspace}
\usepackage{threeparttable}
\usepackage{fullpage}

\usepackage{times}

\usepackage{microtype}
\usepackage{caption}
\usepackage{subcaption}
\DeclareCaptionFont{ls}{\lsstyle} 
\usepackage{xargs}                      

\newcommand{\Atemp}[1]{**\textbf{A Temp: \color{blue}[[[#1]]]}**}
\renewcommand{\Atemp}[1]{}

\newtheorem{theorem}{Theorem}[section]
\newtheorem{lemma}[theorem]{Lemma}
\newtheorem{corollary}[theorem]{Corollary}

\newtheorem{remark}[theorem]{Remark}
\newtheorem{definition}[theorem]{Definition}

\newcommand{\send}{{\bf send~}}
\newcommand{\bdcast}{{\bf broadcast~}}
\newcommand{\terminate}{{\bf Terminate}}
\newcommand{\init}{\hspace{-1em}{\bf Init~}:}
\newcommand{\dbtRound}{\hspace{-1em}{\bf BeginRound~}:}
\newcommand{\finRound}{\hspace{-1em}{\bf EndRound~}:}

\newcommand{\SearchHF}{\textbf{SearchHT}}
\newcommand{\SearchBT}{\textbf{SearchBT}}
\newcommand{\SearchBTDelta}{\textbf{SearchBT\_shift}}

\newcommand{\emptyList}{[\,]}
\newcommand{\addList}{\!\cdot\!}

\newcommand{\Leader}{IsLeader}
\newcommand{\LeaderId}{LeaderId}
\newcommand{\nneigh}{n\_nghbr}
\newcommand{\countBFS}{count}
\newcommand{\parentBFS}{parent}
\newcommand{\pparent}{parent\_port}
\newcommand{\nchild}{n\_child}
\newcommand{\wt}{wt}
\newcommand{\nwt}{n\_wt}
\newcommand{\heavy}{IsHeavy}

\newcommand{\listH}{H}
\newcommand{\listP}{P}
\newcommand{\ivar}{i}
\newcommand{\jvar}{j}
\newcommand{\fport}{fst\_port}
\newcommand{\nport}{nxt\_port}
\newcommand{\listpos}{i\_child}
\newcommand{\last}{last}
\newcommand{\nparent}{nxtparent}

\newcommand{\nparenth}{nxthparent}
\newcommand{\nlchild}{nxtchild_l}
\newcommand{\nrchild}{nxtchild_r}
\newcommand{\heirRT}{heir}
\newcommand{\listNode}{Node}
\newcommand{\listWill}{Will}
\newcommand{\current}{current}
\newcommand{\ncurrent}{nxtPort}
\newcommand{\vark}{k}
\newcommand{\continue}{continue}
\newcommand{\newId}{NewId}
\newcommand{\nextId}{NextId}
\newcommand{\maxId}{Min}
\newcommand{\lightPath}{LightPath}
\newcommand{\routingLabel}{RoutingLabel}

\newcommand{\lpath}{L}

\newcommand{\RT}{\mathrm{RT}}
\newcommand{\will}{\mathrm{Will}}
\newcommand{\wills}{\mathrm{Wills}}
\newcommand{\willportion}{\mathrm{Willportion}}
\newcommand{\willportions}{\mathrm{Willportions}}
\newcommand{\cft}{\mathrm{CompactFT}}
\newcommand{\tz}{\mathrm{TZ}}
\newcommand{\tzft}{\mathrm{CompactFTZ}}

\newcommand{\execute}{\mathrm{execute}}
\newcommand{\executes}{\mathrm{executes}}
\newcommand{\heir}{\emph{heir}}
\newcommand{\helper}{\emph{helper}}
\newcommand{\real}{\emph{real}}

\newcommand{\leafwillportions}{\mathrm{Leafwillportions}}
\newcommand{\leafheir}{\emph{leafheir}}

\newcommand{\CModel}{\mathrm{Compact \, Message \, Passing}}
\newcommand{\CModelAbbrev}{\mathrm{CMP}}

\newcommand{\nonadv}{deterministic}
\newcommand{\adv}{adversarial}

\DeclareGraphicsExtensions{.pdf}

\renewcommand{\algorithmiccomment}[1]{\hfill// #1}

\newcommand{\interval}[2]{\left[{#1},{#2}\right]}

\title{Some Problems in Compact Message Passing}

\author{
Armando Casta\~neda\\ Instituto de Matem\'aticas, UNAM, M\'exico
\and Jonas Lef\`evre \\ Computer Science, Loughborough University, UK
\and Amitabh Trehan \thanks{Corresponding Author. Telephone: +447466670830 (Cell), email: a.trehan@lboro.ac.uk. J. Lefevre and A. Trehan were supported in this research by the EPSRC first grant ref: EP/P021247/1: Compact Self-Healing and Routing Over Low Memory Nodes (COSHER)} \\Computer Science, Loughborough University, UK
}

\date{}

\begin{document}

\maketitle

\begin{abstract}

This paper seeks to address the question of designing distributed algorithms for the setting of \emph{compact memory} i.e. sublinear (in $n$ -- the number of nodes) bits working memory for connected networks of arbitrary topologies. The nodes in our networks may have much lower internal (working) memory (say, $O(poly\log n)$) as compared to the number of their possible neighbours ($O(n)$) implying that a node may not be even able to store the $ID$s of all of its neighbours. These algorithms may be useful for large networks of small devices such as the Internet of Things, for wireless or ad-hoc networks, and, in general, as memory efficient algorithms.

More formally, we introduce the \emph{Compact Message Passing} ($CMP$) model -- an extension of the standard message passing model considered at a finer granularity where a node can interleave reads and writes with internal computations, using a port only once in a synchronous round. The interleaving is required for meaningful computations due to the low memory requirement and is akin to a distributed network with nodes executing streaming algorithms.
Note that the internal memory size upper bounds the message sizes and hence e.g. for $O(\log n)$ memory, the model is weaker than the \textsc{Congest} model; for such models our algorithms will work directly too.

We present some early results in the CMP model for nodes with $O(\log^2 n)$ bits working memory. We introduce the concepts of  \emph{local compact functions} and \emph{compact protocols} and give  solutions for some classic distributed problems (leader election, tree constructions and traversals).  We build on these to solve the open problem of compact preprocessing for the compact self-healing routing algorithm \emph{CompactFTZ} posed in \emph{Compact Routing Messages in Self-Healing Trees} (Theoretical Computer Science 2017) by designing \emph{local compact functions} for finding particular subtrees of labeled binary trees. Hence, we introduce the first fully compact self-healing routing algorithm.  In the process, we also give independent fully compact versions of the Forgiving Tree [PODC 2008] and Thorup-Zwick's tree based compact routing [SPAA 2001].

\end{abstract}


\
\section{Introduction}

 Large networks of low memory devices such as
the \emph{Internet of Things (IOT)} are expected to introduce billions of very weak devices
 that will need to solve distributed computing problems to function effectively.
In this paper, we  attempt to formalise the development of
 distributed algorithms for such a scenario of large networks of low memory devices. We decouple the internal working memory of a node from the memory used by ports for ingress (receiving) and egress (transmitting) (e.g. the ingress queue (Rx) and the egress queue (Tx)) which cannot be used for computation. Thus, in an arbitrary network of $n$ nodes, nodes with smaller internal memory ($o(n)$ bits) may need to support a larger number of connections ($O(n)$).
 To enable this, we introduce the $\CModel$ ($\CModelAbbrev$) model,
the standard synchronous message-passing model at a finer granularity where each process can interleave reads from and writes to its ports
with internal computation using its low ($o(n)$ bits) memory.
We give the first algorithms for several
classic problems in the $\CModelAbbrev$ model, such as leader election (by flooding), DFS and BFS spanning tree construction and traversals and  convergecast. We build on these to develop the first fully compact distributed  routing, self-healing and self-healing compact routing algorithms.
We notice that with careful construction, some (but not all) solutions incurred almost no additional time/messages overhead compared to regular memory message passing.

There has been intense interest in designing efficient routing schemes for distributed networks~\cite{santoro1985labelling, PelegU88, AwerbuchBLP90, Cowen01, ThorupZ01, EilamGP-CRS-2003, GavoilleP-DC2003, AbrahamGM-Routing-2004,Chechik13} with compact routing trading stretch (factor increase in routing length) for memory used.
In essence, the challenge is to use $o(n)$ bits  memory per node overcoming the need for large routing tables or/and packet headers. Our present setting has a similar
ambition - \emph{what all can be done if nodes have limited working memory even if they may have large neighbourhoods?}
In fact, we define local memory as \emph{compact} if it is $o(n)$ bits and by extension, an algorithm as compact if it works in compact memory.

We see two major directions for extending the previously mentioned works.
Firstly, a routing scheme consists of two parts - a pre-processing algorithm (scheme construction) and a routing protocol~\cite{FraigniaudG-STACS-Space02}. The routing results mentioned above assume sequential centralized pre-processing.
Since routing is inherently a distributed networks problem, it makes sense to have the scheme construction distributed too, and this has led to a recent spurt in designing efficient preprocessing algorithms for compact routing schemes~\cite{LenzenP13, GavoilleGHI13, LenzenP15, ElkinN16a, ElkinN17}.
These algorithms do not have explicit constraints on internal working memory, therefore, in essence, they choose to conserve space (for other purposes).
Our interpretation is stricter and we develop a  pre-processing scheme (for a routing scheme from~\cite{ThorupZ01}) assuming that nodes do not even have any excess space and therefore, have, to develop the whole solution in compact memory itself.
Moreover, our solutions are deterministic unlike the solutions listed above, though they tackle a broader range of routing schemes than we do.

Secondly, deterministic routing schemes, in the preprocessing phase, rely on discovery and efficient distributed `encoding' of the network's topology to reduce the memory requirement (a routing scheme on an arbitrary network with no prior topology or direction knowledge would essentially imply large memory requirements).
This makes them sensitive to any topology change and, hence, it is challenging to design fault tolerant compact routing schemes. There has been some work in this direction e.g. in the dynamic tree model ~\cite{KormanPR-TCS-Label04, Korman-DC-CompactLabel07} or with additional capacity and rerouting in anticipation of failures~\cite{ChechikLPR12, Chechik13InfComp, CourcelleT07,doverspike94capacity, FengH01, frisanco97capacity}.
Self-healing is a responsive fault-tolerace paradigm seeking minimal anticipatory additional capacity and has led to a series of work~\cite{Amitabh-2010-PhdThesis, SaiaTrehanIPDPS08, HayesPODC08, Amitabh-SelfHeal-arxiv, Pandurangan2014-DEX, PanduranganXhealT14, FG-DCJournal2012, SarmaTrehanEdgeNetSciComm2012}  in the recent past for maintaining topological properties (connectivity, degrees, diameter/stretch, expansion etc.).
Algorithms were also proposed to `self-heal' computations e.g. \cite{SaadS-DC-17}.
Combining the above motivations, ~\cite{CDT2018-CompactFTZ} introduced a fault-tolerant compact routing solution $\tzft$ in the (deletion only) self-healing model where an omniscient adversary attacks by removing nodes and the affected nodes distributively respond by adding back connections.
However, as in previous routing schemes, $\tzft$'s pre-processing assumed large (not compact) memory.
This paper addresses that important problem developing a compact pre-processing  deterministic algorithm for $\tzft$.
We also develop a compact pre-processing deterministic algorithm for $\cft$ (a compact version of \emph{ForgivingTree}~\cite{HayesPODC08}).
This leads to a fully compact (i.e. completely distributed and in compact memory) routing scheme, a fully compact self-healing routing scheme and a fully compact self-healing algorithm.
 \subparagraph{Our model:}
In brief (detailed model in Section~\ref{sec: model}), our network is an arbitrary connected graph over $n$ nodes. Each node has a number of uniquely identified communication ports.
Nodes have $o(n)$ bits of working memory (We need only $O(\log^2 n)$ for our algorithms).
However, a node may have $\Omega(n)$ neighbours. Note that a node has enough ports for unicast communication with neighbours but port memory is specialised for communication and cannot be used for computation or as storage space.
Also note that the size of the messages are upper bounded by the memory size (in fact, we only need $O(\log n)$ bits sized messages as in the CONGEST model~\cite{peleg}).
 In standard synchronous message passing, in every round, a node reads the messages of all its neighbours, does some internal computation and then outputs messages. Our nodes cannot copy all the messages to the working space, hence, in our model, nodes interleave processing with reads and writes as long as each port is read from and written to at most once in a round.
 Hence, a round may look like $pr_1pr_3pw_2pw_1pr_2 \ldots$ where $p$,$r$ and $w$ stand for processing, reading and writing (subscripted by port numbers). As in regular message passing, all outgoing messages are sent to the neighbours to be delivered by the end of the present round.
 The order of reads may be determined by the node ((self) \nonadv\ reads) or by an adversary (\adv\ reads) and the order of writes by the node itself. We call this the  $\CModel$($\CModelAbbrev$) model.

 The model can also be viewed as a network of machines with each node locally executing a kind of streaming algorithm where the input (of at most $\delta$ items, where $\delta$ is the number of ports) is `streamed' with each item seen at most once and the node computing/outputing results with the partial information.
Our self-healing algorithms are in the \emph{bounded memory deletion only self-healing model} ~\cite{CDT2018-CompactFTZ,Amitabh-2010-PhdThesis} where nodes have compact memory and in every round, an omniscient adversary removes a node but the nearby nodes react by adding connections to self-heal the network. However, their preprocessing requires only the $\CModelAbbrev$ model.
The detailed model is given in Section~\ref{sec: model}.

\begin{figure}[ht!]
 \centering
     \includegraphics[height=3cm]{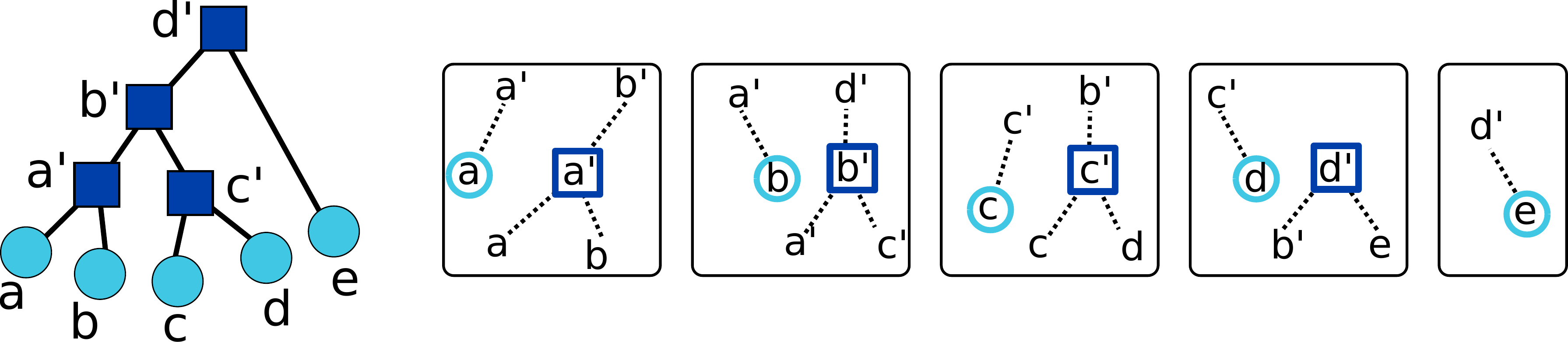}
    \caption{\emph{Compact function $f$ to query labeled BST trees/half-full trees (Section~\ref{subsection-computing-labels})}: On the left is such a tree with 5 leaves.  $f(5,b)$ will return the second box (having the $O(\log n)$ sized subtrees of $b$ and $b'$)  }
        \label{fig: tree and parts}
\end{figure}

\subparagraph{General solution strategy and an example:}
A general solution strategy in the $\CModelAbbrev$  model is to view the algorithms as addressing  two distinct (but related) challenges.
The first is that the processing after each read is constrained to be a function of the (memory limited) node state and the previous read (as in a streaming or online fashion) and it results in an output (possibly NULL) which may be stored or output as an outgoing message.
We will refer to such a function as a \emph{local compact function}. The second part is to design, what we call, a \emph{compact protocol} that solves the distributed problem of passing messages to efficiently solve the chosen problem in our model. We discuss local compact functions a bit further in our context. A simple compact function may be the $max(.)$ function which simply outputs the maximum value seen so far. A more challenging problem is to design a function that outputs the neighbourhood of a node in a labelled binary tree. Consider the following question: \emph{Give a compact function that given as input the number of leaves $n$ and any leaf node $v$, returns the neighbourhood of $v$ for a balanced binary search tree of $n$ leaves with internal nodes repeated as leaves and arranged in ascending order from left to right}. Note that the function should work for a tree of any size without generating the whole tree (due to limited memory). Figure~\ref{fig: tree and parts} illustrates the question (further background is in Section~\ref{sec: cftz overview}) -- the solution to a similar question (solution in Section~\ref{subsection-computing-labels}) forms the crux of our fully compact algorithms. It's also a question of interest whether this approach could be generalised to construct a generic function that when input a compact description of a structure (in our case, already encoded in the function) generates relevant compact substructures on demand when queried.

\subparagraph{Our results:}
Our results follow. We introduce the model $\CModelAbbrev$ hoping it will provide a formal basis for designing algorithms for devices with low working memory in large scale networks.
 As mentioned, we introduce a generic solution strategy (compact protocols with compact functions) and in Section~\ref{subsection-computing-labels} (cf. Lemma~\ref{lm: searchhf}), we give a compact function  of independent interest that compactly queries a labelled binary tree.
We give some deterministic algorithms in the $\CModelAbbrev$ model as summarised in Table ~\ref{tab: results}. We do not provide any non-obvious lower bounds but for some algorithms it is easy to see that the solutions are optimal and suffer no overhead due to the lower memory as compared to regular message passing (denoted with a `*' in Table~\ref{tab: results}).  In general, it is easier to construct (by broadcast) and maintain spanning trees using a constant number of parent and sibling pointers, and effectively do bottom up computation, but unicast communication of parent with children may suffer due to the parent's lack of memory (with parent possibly resorting to broadcast). We solve preprocessing for the compact routing scheme $TZ$ (\cite{CDT2018-CompactFTZ}, based on~\cite{ThorupZ01}), compact self-healing scheme $\cft$ (\cite{CDT2018-CompactFTZ}, based on~\cite{HayesPODC08}) and $\tzft$ as summarised in Theorem \ref{th: main} leading to fully compact routing, self-healing and self-healing routing solutions (in conjunction with~\cite{CDT2018-CompactFTZ}) as Corollaries~\ref{cor: ftz} to ~\ref{cor: ftzft}.
Note that combining with the tree cover results from~\cite{Thorup2005},  our algorithms could be extended to obtain low stretch routing for general graphs.
 In Section~\ref{subsection-computing-labels}, we also give a compact function for outputting the neighbourhood of a node of a labelled binary tree (too big to fit into the memory) which may be of independent interest (Lemma~\ref{lm: searchhf}).


\begin{threeparttable}[tbh]
\begin{tabular}{|lcccr|}
  \hline
  Algorithm    & Internal  & Time & \#\,messages & In Paper\\
            & memory &  (\#\,rounds) &  & \\
  \hline
$\star\,$Leader Election by flooding    &   $O(\log n)$   &  $O(D)$         &    $O(m)$         & Section~\ref{sec: floodLE}\\
$\star\,$BFS Spanning Tree Construction           &   $O(\log n)$   &  $O(D)$         &    $O(m)$         & Section~\ref{sec:bfs}\\
$\star\,$(regular) Convergecast &  $O(\log n)$    &   $O(D)$        &    $O(n)$       &Section~\ref{sec:cvgcast} \\
Weight labelling with convergecast   &  $O(\log n)$    &   $O(D)$        &    $O(nD)$     & Section~\ref{sec:cvgcast}\\
Compact DFS relabelling of a tree  &   $O(\log n)$   &   $O(m)$        &    $O(m)$          &Section~\ref{sec:dfs}\\
$\star\,$further Compact DFS walks &   $O(\log n)$   &   $O(n)$        &    $O(n)$          &Section~\ref{sec:dfs}\\
`Light' Path by BFS construction &   $O(\log^2 n)$   &   $O(D)$        &    $O(m)$       &  Section~\ref{sec:dfs} \\
\hline
Wills (w/ half-full tree labelling) setup & & & &\\
$\star\,$ \dots with \nonadv\ reads &   $O(\log^2 n)$ & $O(1)$  & $O(n)$      &Section~\ref{subsection-computing-will}\\
\phantom{$\star\,$} \dots with \adv\ reads   &   $O(\log n)$ & $O(\Delta)$ & $O(n\Delta)$    & Section~\ref{subsection-computing-will-adv} \\
  \hline
 $\tz$ preprocessing & $O(\log^2 n)$  & $O(m)$  &$O(m)$  & Th.\ref{th: main}.\ref{subth: tz}\\
 $\cft$ preproc. with \adv\ reads & $O(\log n)$ & $O(D\!+\!\Delta)$ & $O(m\!+\!n\Delta)$ 	&	Th.\ref{th: main}.\ref{subth: cft}\\
 \phantom{$\cft$ preproc.} with \nonadv\ reads & $O(\log n)$ & $O(D)$ & $O(m)$ 				&	Th.\ref{th: main}.\ref{subth: cft}\\
 $\tzft$ preproc. with \adv\ reads & $O(\log^2 n)$ & $O(m\!+\!\Delta)$ & $O(m\!+\!n\Delta)$ &	Th.\ref{th: main}.\ref{subth: tzft}\\
 \phantom{$\tzft$ preproc.} with \nonadv\ reads & $O(\log^2 n)$ & $O(m)$ & $O(m)$  			&	Th.\ref{th: main}.\ref{subth: tzft-log}\\
 \hline
\end{tabular}
 \begin{tablenotes}
 \item[$\dagger$] $\Delta$: Upper bound on number of ports of a node
 \item[$\star$] No additional overhead in comparison with regular message passing.
\end{tablenotes}
\caption{Summary of the algorithms in the CMP model in this paper.
Results apply to both \nonadv\ and \adv\ reads unless otherwise indicated.
\label{tab: results} }
\end{threeparttable}

\begin{theorem}
\label{th: main}
In the $\CModel$ model, given a connected synchronous network $G$ of $n$ nodes and $m$ edges with $O(\log^2 n)$ bits local memory:
\begin{enumerate}
\item \label{subth: tz}  Algorithm $\tz$ preprocessing (\nonadv\ / \adv\ reads) can be done using $O(\log n)$ bits size messages  in $O(m)$ time and $O(mD)$ messages, or using $O(\log^2 n)$ bits size messages in $O(m)$ time and $O(m)$ messages.
\item \label{subth: cft} Algorithm $\cft$ preprocessing can be done using  $O(\log n)$
 bits size messages in $O(D)$ time and $O(m)$ messages for  \nonadv\  reads, or $O(D + \Delta)$ time and $O(m + n\Delta)$ messages  for \adv\ reads.
\item \label{subth: tzft-log}
Algorithm $\tzft$ preprocessing can be done using $O(\log n)$ bits size messages in $O(m)$ time and $O(mD)$ messages for \nonadv\ reads or $O(m + \Delta)$ time and $O(mD +n\Delta)$ messages for  \adv\ reads.
\item \label{subth: tzft} Algorithm $\tzft$ preprocessing can be done using  $O(\log^2 n)$ bits size messages in $O(m)$ time and $O(m)$  messages for  \nonadv\ reads, or $O(m + \Delta)$ time and $O(m + n\Delta)$ messages for \adv\ reads.
\end{enumerate}
where $\Delta$ is an upper bound on the number of ports of a node
\end{theorem}

\begin{corollary}
\label{cor: ftz}
There is a fully compact and distributed variant of the routing scheme $\tz$ for a network of nodes
with $O(\log^2 n)$ bits memory, $O(\log n)$ bits routing tables, and $O(\log^2 n)$ bits labels and message size.
\end{corollary}

\begin{corollary}
\label{cor: fcft}
There is a fully compact and distributed variant of the self-healing algorithm $\cft$ for a network of nodes with $O(\log n)$ bits memory and message size.
\end{corollary}

\begin{corollary}
\label{cor: ftzft}
There is a fully compact and distributed variant of the self-healing routing algorithm $\tzft$ for a network of nodes with $O(\log^2 n)$ bits internal memory.
\end{corollary}


\subsection{Warm up: Leader Election by Flooding}
\label{sec: floodLE}

 As a warm up, let us implement flooding and use it to elect a leader. Assume that a node $v$ has a message $M$ in its memory that needs to be flooded on the network. Node $v$  executes the primitive \emph{Broadcast} $M$ (Sec.~\ref{sec: modeldetailed}):  $v$ sweeps through its ports in order copying $M$ to every port to be sent out in the next round. In the next round, all nodes, in particular, the neighbours of $v$ read through their ports in \nonadv\ or \adv\ order and receive $M$ from $v$.
  $M$  is copied to the main memory
   and subsequently broadcast further.
   To adapt the flooding algorithm for leader election, assume for simplicity that all nodes wake up simultaneously, have knowledge of diameter ($D$) and elect the highest $ID$ as leader. Since every node is a contender, it will broadcast its own ID: say, $v$ broadcasts $M_v$ (message with $ID$ $v$) in the first round. In the next round, every node will receive a different message from its neighbours.

\renewcommand{\algorithmicrequire}{\textbf{Receive }}

\begin{algorithm}[htb]
\caption{Leader Election By Flooding Rules :}
\label{alg:Leader}
\footnotesize

\begin{tabular}{p{0.45\linewidth}p{0.45\linewidth}}
\begin{algorithmic}
	\STATE{\init}
	   \STATE{ $\Leader \leftarrow \TRUE$}
	   \STATE{$\LeaderId \leftarrow MyId$}
	   \STATE{\bdcast $< LEADER, myId >$}
\end{algorithmic}

\bigskip

\begin{algorithmic}
\STATE{\dbtRound}
	   \STATE{$update\leftarrow \FALSE$}
\end{algorithmic}

&


\begin{algorithmic}
\REQUIRE{$< LEADER,Id >$ from port $X$:}
	\IF{ $Id > \LeaderId$}
	   \STATE{ $\Leader \leftarrow \FALSE$}
	   \STATE{$\LeaderId \leftarrow Id$}
	   \STATE{$update\leftarrow \TRUE$}
	\ENDIF
\end{algorithmic}

\medskip

\begin{algorithmic}
\STATE{\finRound}
	\IF{ $update$}
	   \STATE{ \bdcast $< LEADER,\LeaderId >$}
	\ENDIF
\end{algorithmic}
\\

\end{tabular}

\end{algorithm}

Since a node may have a large number of neighbours, it cannot copy all these $ID$s to the main memory (as in standard message passing) and deduce the maximum. Instead, it will use the interleaved processing in a streaming/online manner to find the maximum ID received in that round. Assume that a node $v$ has a few neighbours \{$a, b, d, f, \ldots $ \} and the reads are executed in order $r_b, r_d, r_a, \ldots$ and so on. To discover the maximum $ID$ received, $v$ simply compares the new $ID$ read against the highest it has: let us call this function $\max$ (this is a \emph{locally compact} function). Therefore, $v$ now executes in an interleaved manner  $r_b \max\, r_d \max\, r_a \max \ldots$. At the end of the round, $v$ has the maximum $ID$ seen so far. Every node executes this algorithm for $D$ synchronous rounds to terminate with the leader decided.
Note the algorithm can be adapted to other scenarios such as non-simultaneous wakeup and knowledge of $n$ (not $D$) with larger messages or more rounds.
Without knowledge of bounds of $n$ or $D$, an algorithm such as in~\protect\cite{KuttenPP0T-JACM-LE-15} (Algorithm 2) can be adapted (not discussed in this paper). {The pseudocode given as Algorithm~\ref{alg:Leader} leads to the local variable $\LeaderId$ being set to the leader's ID.  The local variables of a node used in our algorithms are given in table~\ref{tab: algovariables} along with the stage at which they are set to the right value.}


\section{Model}
\label{sec: model}

We assume a connected network of arbitrary topology represented by an undirected graph $G=(V,E)$ with $|V|=n$ and $|E|=m$ for $n$ nodes and $m$ bidirectional links.
Every node has \emph{compact} internal memory (of size $k = o(n)$),  a \emph{unique id} and a collection of \emph{ports} (interfacing with the bidirectional links) each with a a locally unique \emph{port-id}.
Each port has an \emph{in-buffer} that can be read from and an \emph{out-buffer} that can be written to.
Note that the ports need not be physical devices but even uniquely identifiable virtual interfaces e.g. unique frequencies in wireless or sensor networks.
Also, the neighbours need not be on contiguous ports i.e. there may be `dead' ports interspersed with live ones.
This may happen, for example, when even though starting from contiguous ports, certain neighbours get deleted (as in our self-healing scenarios) or subnetworks (e.g. spanning trees) are generated from the original network.
Therefore, our algorithms have to be aware of such `dead' ports.
For this work, we assume a synchronous network i.e. the communication between the nodes proceeds in synchronous rounds.

\subsection{Compact Memory Passing model}
\label{sec: modeldetailed}

{In this work, we are interested in overcoming the assumption of unlimited computation power of nodes by restricting their internal memory.
This is a natural condition for many real life networks such as the increasingly prevalent networks of low memory devices as in the Internet of Things (IOT), and for applications such as compact routing (which limit memory usage).
The main criteria is to limit node memory to $o(n)$. We do not ask for a bound on the degree of nodes in the network.
This implies that a node may not be even able to store the IDs of all its neighbours in its internal memory if it has $o(n)$ memory.
A parametrised version would be a \emph{(B,S)-compact} model where $B$ is the local memory size and $S$ is the maximum size of a message exchanged at each edge.
For example, we could be interested in a sublinear CONGEST model with $B=O(\log n)$ and $S=O(\log n)$.
Notice that $S$ can not exceed $B$ since a node needs to generate the message in its internal memory before sending it.
The case of $(O(\log^2 n), O(\log n))$-compact would be naturally interesting since this would be comparable with the standard CONGEST model (with low internal memory) while allowing applications such as compact routing.
Since a node might not be able to store all the information sent to it in a round of communication, to allow meaningful computation in compact memory, we need to revisit the standard message passing model (at a finer granularity).
Hence, we introduce the \emph{$\CModel$ ($\CModelAbbrev$)} model and its variants.
}

In the standard synchronous message passing model, nodes are assumed to have unlimited computation.
In each round, a node reads all its ports, possibly copying  received messages to internal memory, processing the inputs preparing messages which are then written to the ports for transmission.
However, our nodes having compact memory cannot store the inputs (or even $ID$s) of their neighbours
Hence, we propose the following model with a streaming style computation.

\paragraph{$\mathbf{\CModel}$ ($\mathbf \CModelAbbrev$):}
Communication proceeds in synchronous rounds.
Internally, in every round, every node $v$ executes a sweep of its ports fulfilling the following conditions:
\begin{enumerate}
\item~\label{cond: mutable} \emph{Mutable reads condition:} If a read is executed on an in-buffer, the value in that buffer is cleared i.e. not readable if read again.
\item~\label{cond: fairness} \emph{Fair interleaving condition:} In a sweep, $v$ can read and write to its ports in any order interleaving the reads and writes with internal processing
i.e.  $pr_ipw_{i'}r_jpw_{j'}p \ldots$, where $p$,$r$ and $w$ stand for processing (possibly none), reading and writing (subscripted by port numbers ( $i,i',j,j',\ldots$).
For example, $pr_1pr_2pw_2pw_1p \ldots$ are $pr_1pw_2pr_3pw_1\ldots$ are  valid orders.
Note that the memory restriction bounds the local computation
between reads and writes in the same round.
We say that such computations are given by \emph{locally compact functions} where a locally compact function takes as input the previous read and the node state to produce the next state and message(s).
{e.g. in the extreme case constant local memory allows only a constant the local computation between reads and writes is constant too.}
\begin{enumerate}
\item \emph{(self) \nonadv\ reads:} $v$ chooses the order of ports to be read and written to provided that in a sweep, a port is read from and written to at most once. Note that the node can adaptively compute the next read based on previous reads in that sweep.
\item \emph{\adv\ reads:}
An adversary decides the order of reads i.e. it picks one of the possible permutations over all ports of the node.
The order of writes is still determined by the node $v$ itself. A special case is a \emph{randomized adversary} that chooses a random permutation of reads. A \emph{strong adversary} can adaptively choose the permutation depending on the node state.
\end{enumerate}
\end{enumerate}

We define the following primitives: ``{\bf Receive~} $M$ from $P$"  reads the message $M$ from in-buffer $P$ to the internal memory; ``{\bf Send~} $M$ via $P$" will write the message $M$ to the out-buffer of port $P$
and ``{\bf Broadcast~}$M$'' will write the message $M$ on every port of the node for transmission.
Since condition~~\ref{cond: fairness} limits writes to one per port, we also define a primitive
``{\bf Broadcast~}$M$ except $listP$'' which  will `send' the message on each port except the ones listed in $listP$.
This can be implemented as a series of Sends where the node checks $listP$ before sending the message. Notice that $listP$ has to be either small enough to fit in memory or a predicate which can be easily computed and checked. For ease of writing, we will often write the above primitives in text in a more informal manner in regular English usage i.e. receive, send, broadcast, and `broadcast except to ...' where there is no ambiguity. A message is of the form $<\mathrm{Name\ of\ the\ message},\mathrm{Parameters\ of\ the\  message}>$.

{\begin{figure}[bt!]
\centering
\includegraphics[height=3cm]{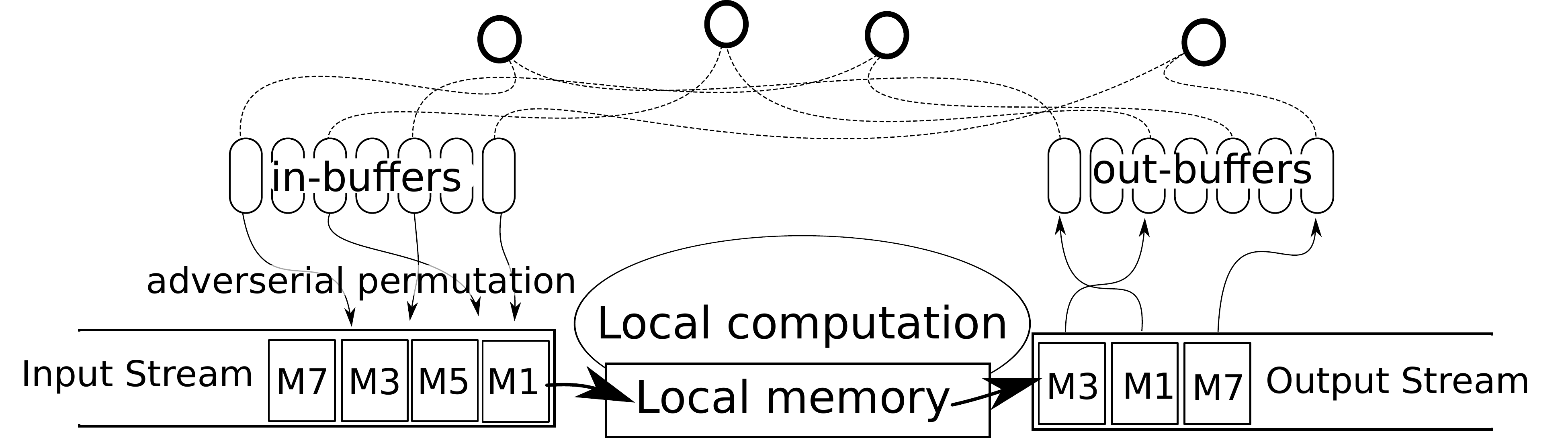}
\caption{The $\CModelAbbrev$ model can be viewed as each node executing a kind of streaming algorithm with read-only input and write-only output tapes specifying connections to neighbours.}
\label{fig: nodeTM}
\end{figure}
}

The model can be viewed as each node executing some kind of streaming algorithm (ref. Figure~\ref{fig: nodeTM}).
The incoming messages form a continuous stream in a permutation decided by an adversary (or the algorithm itself).
The internal algorithm reads this stream and produces, using its restricted local memory, a stream of outcoming messages that fills the out-buffers.

\pagebreak
\section{Background (CompactFTZ) and paper layout}
 \label{sec: cftz overview}

\newcommand{\myalgorithm}
{
\small{
\flushleft \emph{$\tzft$: Compact Preprocessing followed by Compact self-healing routing}
\begin{algorithmic}[1]
\STATE \label{pre: le}  Given a distinguished node $v$ (e.g. by compact leader election ~(Sec.~\ref{sec: floodLE}))
\STATE \label{pre: bfs}  $T_a \leftarrow$  A BFS spanning tree of graph $G_0$~(Sec.~\ref{sec:bfs})
\STATE \label{pre: cc} $T_b \leftarrow$ Setup of \emph{$\tz$ heavy arrays} by \emph{compact convergecast} of $T_a$~(Sec.~\ref{sec:cvgcast})
\STATE  \label{pre: dfs} $T_c \leftarrow$ DFS traversal and labelling (renaming) of $T_b$~(Sec.~\ref{sec:dfs})
\STATE  \label{pre: llev}$T_d \leftarrow$ Setup of \emph{$\tz$ light levels} by BFS \emph{traversal} of $T_c$~(Sec.~\ref{subsec:prepoc})\!
\STATE  \label{pre: wills} $T_e  \leftarrow $ Setup of \emph{$\cft$ $\wills$} ~(Sec.~\ref{sec:cftree})
\WHILE {true}
\IF[\emph{$\cft$ Self-healing~\cite{CDT2018-CompactFTZ}}] {a vertex $x$ (with parent $p$) is deleted} \label{algoline: deletion}
\IF[Fix non leaf deletion]  {$x$ was not a leaf (i.e., had any children)}
\STATE $x$'s children $\execute$ $x$'s Will using $x$'s $\willportions$ they have; $x$'s $\heir$ takes over $x$'s duties. \label{algoline: fixnonleafstart}
\STATE All affected Wills are updated by simple update of relevant $\willportions$.  \label{algoline: fixnonleafend}
\ELSE[Fix leaf deletion]  \label{algoline: fixleafstart}
\IF[Update Wills by simulating the deletion of $p$ and $x$]  {$p$ is $\real$/alive}
\STATE $p$ informs children about deletion; they update $\leafwillportions$ exchanging messages via $p$
\ELSE[$p$ had already been deleted earlier] \label{algoline: fixnonleafvp}
\STATE \label{algoline: fixnonleafvpend} Let $y$ be $x$'s $\leafheir$; $y$ $\executes$ $x$'s $\will$ and affected nodes update $\willportions$.
\ENDIF
\ENDIF
\ENDIF
\IF[\emph{Compact Self-Healing Routing}]{A message headed for $w$ is received at node $v$} \label{algoline: msgarrival}
\IF[Deliver over regular network via compact routing scheme $\tz$]{$v$ is a real node}
\STATE If ($v = w$) message has reached else if $w \notin [d_v, v]$ forward to parent else if $w \in [c_v,v]$ forward to \emph{light} node through port $L(w)[\ell_v]$ else forward to a \emph{heavy node} through port $P_v[i]$
\ELSE[$v$ is a virtual helper node (= $\helper(v))$]
\STATE  If ($v = w$) message has reached else traverse $\RT$ in a binary search manner
\ENDIF
\ENDIF
\ENDWHILE
\end{algorithmic}
}
}

\begin{algorithm}[bht]
    \centering
        \myalgorithm
        \caption{
       $\tzft$ with  preprocessing: A high level view}
                \label{algo: tzftpreproc}
\end{algorithm}


Here, we give a brief background of $\tz$, $\cft$ and $\tzft$ referring to the relevant sections for our solutions. Note that some proofs and pseudocodes have been omitted from the paper due to the lack of space.
Algorithm~\ref{algo: tzftpreproc} captures essential details of $\tzft$ (and of $\tz$ and $\cft$). These algorithms, like most we referred to in this paper, have a distinct preprocessing and main (running) phase. The data structures are setup in the preprocessing phase to respond to events in the main phase (node deletion or message delivery). First, let us consider the intuitive approach.  $\tzft$ is designed to deliver messages between sender and receiver (if it hasn't been adversarially deleted) despite node failure (which is handled by self-healing). Self-healing ($\cft$) works by adding virtual nodes and new connections in response to deletions. Virtual nodes are simply logical nodes simulated in real (existing) nodes' memories. Thus, the network over time is a patchwork of virtual and real nodes. It is now possible (and indeed true in our case) that the routing scheme $\tz$ may not work over the patched (self-healed) network and the network information may be outdated due to the changes. Thus, the composition $\tzft$ has two distinct routing schemes and has to ensure smooth delivery despite outdated information.
Nodes then respond to the following events: i) \emph{node deletion (line~\ref{algoline: deletion}):} self-heal using $\cft$ moving from initial graph $G_0$ to $G_1$ and so on (the $i^{th}$ deletion yielding $G_i$), or ii) \emph{message arrival (line~\ref{algoline: msgarrival}):} Messages are forwarded using  $\tz$ or the second scheme (which is simply binary search tree traversal in our case).

\begin{figure}[h!]
 \centering
     \includegraphics[height=3cm]{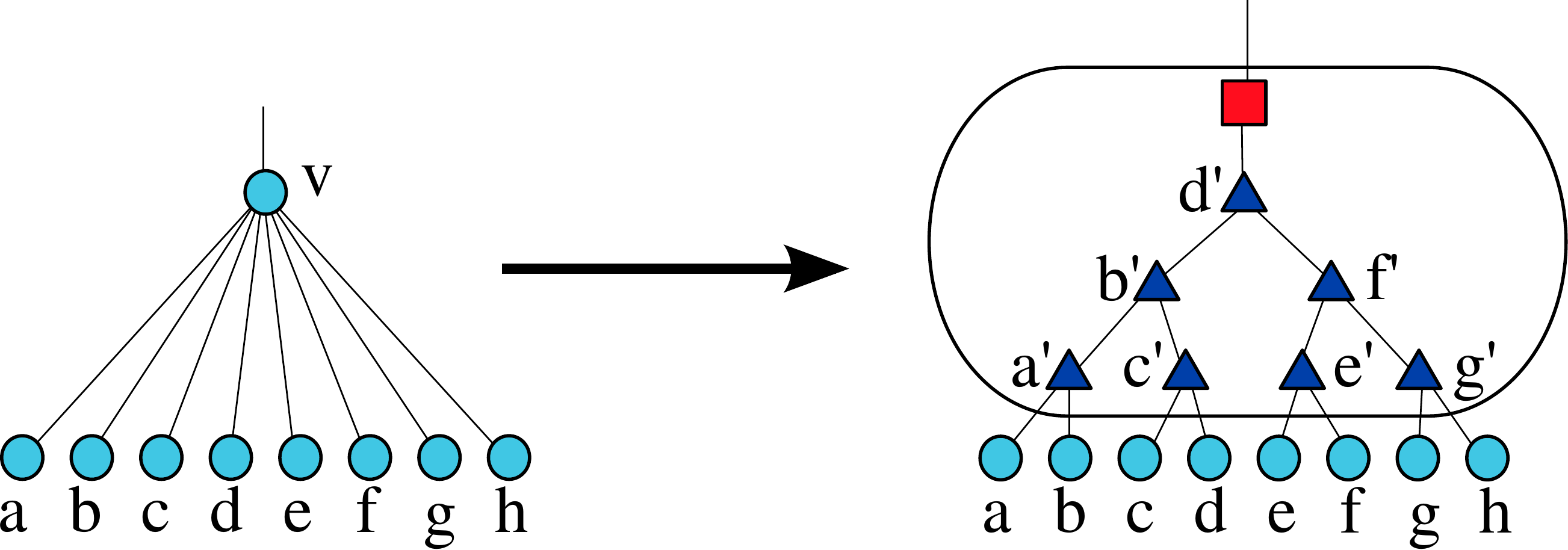}
    \caption{$\cft$~\cite{CDT2018-CompactFTZ, HayesPODC08}: The deleted node $v$ is replaced by a reconstruction tree ($\RT$) with $v$'s ex-neighbours forming the leaves and simulating the internal virtual nodes (e.g. $d'$ is simulated by $d$).}
        \label{fig: MakeRTColour}
\end{figure}

Consider $\cft$. $\cft$ seeks to limit diameter increase while allowing only constant (+3) node degree increase over any sequence of node deletions. Starting with a distinguished node (line~\ref{pre: le}), it constructs a  BFS spanning tree in the preprocessing (line~\ref{pre: bfs}) and then sets up the healing structures as follows. A central technique used in topological self-healing  is to replace the 
deleted subgraph
 by a reconstruction subgraph of its former neighbours (and possibly virtual nodes simulated by them). These subgraphs have been from graph families such as balanced binary search trees~\cite{HayesPODC08}, half-full trees~\cite{FG-DCJournal2012}, random r-regular expanders~\cite{PanduranganXhealT14}, and p-cycle deterministic expanders~\cite{Pandurangan2014-DEX}. 
Figure~\ref{fig: MakeRTColour} illustrates this for $\cft$ where the star graph of deleted node $v$ is replaced by the \emph{Reconstruction Tree}($\RT$) of $v$.
 In preprocessing (line~\ref{pre: wills}), every node constructs its $\RT$ (also called its $\will$) in memory and  distributes the relevant portions (called $\willportion$) to its neighbours so that they can form the $\RT$ if it is deleted. However, since nodes do not have enough memory to construct their $\RT$, they rely on a compact function to generate the relevant will portions. Referring back to Figure~\ref{fig: tree and parts}, the tree in the figure can be thought of as a $\RT$ of a deleted node (or its  $\will$ before demise) and the subgraphs in the boxes as the $\willportions$ (one per neighbour). The node now queries the compact function \SearchHF(Algorithm~\ref{alg-search-hf}) to generate $\willportions$. Once these structures have been setup in preprocessing, the main phase consists of `executing' the $\will$ i.e. making the new edges upon deletion and keeping the  $\willportions$ updated. The actions differ for internal and leaf nodes -- cf.~\cite{CDT2018-CompactFTZ} for details.

Now, consider the routing scheme $\tz$. $\tz$ is postorder variant of the tree routing scheme of~\cite{ThorupZ01}. The scheme is wholly constructed in the preprocessing phase - the original paper does not give a distributed construction. Here, we give a compact distributed construction.  On a rooted spanning tree (the BFS tree obtained for $\cft$ above), every node is marked either \emph{heavy} if it heads a subtree of more than a $b^{th}$ (a constant) fraction of its parent's descendants else \emph{light}. Reference to (at most $b$) heavy children is stored in an array $H$ with corresponding ports in an array $P$ making a routing table.  We do this by a \emph{compact convergecast} (Line~\ref{pre: cc}). A DFS traversal prioritised by heavy children follows; nodes now get relabeled by their DFS numbers (line~\ref{pre: dfs}).  Lastly, for every node, its path from the root is traced and the light nodes on the way (which are at most $O(\log n)$) are appended to its new label (line~\ref{pre: llev}). Every node  now gets a `light level' as the number of light nodes in its path. Note that the label is of $O(\log^2 n)$ bits requiring our algorithms to use $O(\log^2 n)$ bits memory. All other parts (including $\cft$) require only  $O(\log n)$ bits. This yields a compact setup of $\tz$. When a packet arrives, a real node checks its parent and array $H$ for the DFS interval failing which it uses its light level and the receiver's label to route the packet through light nodes. If a packet comes to a virtual node,  binary search traversal is used since our $\RT$s are binary search trees.
 Interestingly,  even though the arrays and light levels etc. get outdated due to deletions, \cite{CDT2018-CompactFTZ} shows routing continues correctly once set up.

\section{Some Basic Tree Algorithms and $TZ$ Preprocessing}
\label{sec: basictz}

We present here three distributed algorithms related to trees:
(1) BFS traversal and spanning tree construction,
(2) convergecast and
(3) DFS traversal, tree construction and renaming.
We present these independently and also adapting them in the context of $\tz$, $\cft$ and $\tzft$ preprocessing.
The general algorithms can be easily adapted for other problems, for example, the BFS construction can be adapted to compute compact topdown recursive functions, convergecast for aggregation and bottom-up recursive functions and DFS to develop other priority based algorithms.

\begin{table}[!h]
    \small
\begin{tabular}{|r@{ : }lr|}
  \hline
{$\mathbf{\nneigh}$} & the total number of neighbours &\emph{(already set)}\\
{$\mathbf{\Leader}$} & a boolean to recognise the root of the tree & \emph{(set by Alg.\ref{alg:Leader})}\\
{$\mathbf{\LeaderId}$} & the Id of the root of the tree & \emph{(set by Alg.\ref{alg:Leader})}\\
{$\mathbf{\parentBFS}$ and $\mathbf{\pparent}$} & the ID and port of the parent in the BFS tree&\emph{(set by Alg.\ref{alg:tree})}\\
{$\mathbf{\nchild}$} & the number of node that accepted the node as parent&\emph{(set by Alg.\ref{alg:tree})}\\

{$\mathbf{\wt}$} & the weight of the node&\emph{(set by Alg.\ref{alg:heavy})}\\
{$\mathbf{\heavy}$} & a boolean to say if the node is an heavy child (of its parent)&\emph{(set by Alg.\ref{alg:heavy})} \\

{$\mathbf{\listH}$ and $\mathbf{\listP}$} & the lists of heavy node Id ($\listH$) and port ($\listP$) (both of length $\left|\listH\right|$)&\emph{(set by Alg.\ref{alg:rename})}\\
{$\mathbf{\fport }$} & a port id& \emph{(set by Alg.\ref{alg:rename})}\\
{$\mathbf{\nport }$} & the ID of one port from the parent&\emph{(set by Alg.\ref{alg:rename})}\\
{$\mathbf{\listpos }$} & \begin{tabular}{l} an integer representing the position of the node\\  \quad in the list of the children of its parent\end{tabular} &\emph{(set by Alg.\ref{alg:rename})}\\
{$\mathbf{\last }$} & the ID of last used port&\emph{(set by Alg.\ref{alg:rename})}\\
{$\mathbf{\newId }$} & the new ID given by the server& \emph{(set by Alg.\ref{alg:rename})}\\
{$\mathbf{\maxId }$} & the smallest of the new ID among all the node in the subtree &\emph{(set by Alg.\ref{alg:rename})}\\
{$\mathbf{\listNode }$} & list of needed node informations: a counter of uses, the node ID and the port ID     &\emph{(set by Alg.\ref{alg:will})}\\
{$\mathbf{\listWill }$} &  list of partially  computed will parts &\emph{(set by Alg.\ref{alg:will})}\\
\hline
\end{tabular}
\caption{Table of variables of a node (used by the algorithms in Section~\ref{sec: floodLE} and Sections
\ref{sec:bfs} to~\ref{sec:cftree})}
\label{tab: algovariables}
\end{table}

\subsection{Breadth First Traversal and Spanning Tree Construction}
\label{sec:bfs}

\renewcommand{\algorithmicrequire}{\textbf{Receive }}

\begin{algorithm}[htb]
\caption{BFS Tree construction Rules :}
\label{alg:tree}
\footnotesize

\begin{tabular}{p{0.45\linewidth}p{0.45\linewidth}}

			\begin{algorithmic}
	\STATE{\init}
	\IF{ $\Leader$}
	\STATE{\bdcast $<\!JOIN, Id(Root)\!>$}
	\ENDIF
	\end{algorithmic}

\bigskip

	\begin{algorithmic}
	\REQUIRE{{ $<\!YES\!>$ from $X$:}}
		\STATE{$\nchild\!+\!\!+$}
		\IF{$\countBFS+\nchild +1==\nneigh$}
		\STATE{\terminate}
		\ENDIF
		\IF{$\nchild ==\nneigh$ \AND $\Leader$}
		\STATE{\terminate}
		\ENDIF
	\end{algorithmic}

	&

\begin{algorithmic}
\REQUIRE{$<\!JOIN,Id\!>$ from port $X$:}
	\IF{ $\parentBFS==\bot$}
	\STATE{ $\parentBFS \leftarrow Id$ ; $\pparent \leftarrow X$}
	\STATE{ \send $<\!YES\!>$ via $X$}
	\STATE{ \bdcast $<\!JOIN,myId\!>$ except $X$}
	\ELSE
	\STATE{ $\countBFS\!+\!\!+$}
	\ENDIF
	\IF{$\countBFS + \nchild +1==\nneigh$}
	\STATE{\terminate}
	\ENDIF
\end{algorithmic}

\\
\end{tabular}
\end{algorithm}

We assume the existence of a Leader (Section~\ref{sec: floodLE}). Namely each agent has a boolean variable called $isLeader$
such that this variable is set to $False$ for each agent except exactly one.
This Leader will be the root for our tree construction.
The construction follows a classic Breadth First Tree construction.
The root broadcasts a $JOIN$ message to all its neighbours.
When receiving a $JOIN$ message for the first time a node {\it joins} the tree:
it sets its $\parentBFS$ and $\pparent$ variables with the node and port ID from the sender,
it answers $YES$ and broadcasts $JOIN$ further.
It will ignore all next $JOIN$ messages.
To ensure termination, each node counts the number of $JOIN$ and $YES$ messages it has received so far,
terminating the algorithm when the count is equal to the number of its neighbours.

 \pagebreak

\begin{lemma}
After $O(D)$ rounds, every node has set its $\parentBFS$ pointer and reached the state $Terminate$.
\end{lemma}
\begin{proof}
The graph network is connected.
Therefore a path of length at most $D$ exists between the root and any node $v$.
The root is from the start in the tree.
While not all the nodes of the path joined,
each round at least one node joins the tree.
After $D$ rounds all the nodes of the path have joined.

After joining a node sends a message on each link: $YES$ to its parent and $JOIN$ to the others.
The nodes do not send any other message.
This means that each node will receive one message from each link.
Each message increments one counter ($\countBFS$ or $\nchild$) except for the accepted $JOIN$ message.
Eventually the sum of those counters will reach their maximum values and the node will enter $\terminate$ state.
\end{proof}

\begin{lemma}
 There is no cycle in the graph induced by the $\parentBFS$ pointers.
\end{lemma}
\nopagebreak
\begin{proof}
A node sends a message $JOIN$ only if it has already joined the tree.
Then a node accepts a $JOIN$ message only from a node already inside the tree.
If we label the node with the time where they joined the tree, a parent has a time-label strictly smaller than any one of its children.
This implies there can not be a cycle.

An other simpler argument is to notice that we have a connected graph with $n$ vertices and $n-1$ edges (each node has one parent except the root), then it is a tree.
\end{proof}

\begin{corollary}
 Algorithm~\ref{alg:tree} constructs a BFS spanning tree in $O(D)$ rounds using $O(m)=O(n^2)$ messages overall.
\end{corollary}

\subsection{Convergecast}
\label{sec:cvgcast}

\renewcommand{\algorithmicrequire}{\textbf{Receive }}

\begin{algorithm}[!htb]
\caption{Weight Computation by Convergecast - Rules :}
\label{alg:heavy}
\footnotesize

\begin{tabular}{p{0.45\linewidth}p{0.5\linewidth}}

\begin{algorithmic}
	\STATE{\init}
	\IF{$\nchild==0$}
\STATE{ $\wt \leftarrow 1$ ; $\continue \leftarrow \TRUE$}
	\STATE{\send  $<\!WT,1\!>$ via $\parentBFS$}
\ELSE
\STATE{ $\wt \leftarrow 0$ ; $\continue \leftarrow \FALSE$}
	\ENDIF
\end{algorithmic}

\medskip

\begin{algorithmic}
\STATE{\dbtRound}
		\IF{$\continue$}
		\STATE{ \send   $<\!WT2,\wt\!>$ via $\pparent$}
		\ENDIF
\end{algorithmic}

\medskip

\begin{algorithmic}
\REQUIRE{\send   $<\!WT',h\!>$ from $X$:}
\STATE{ $\heavy \leftarrow h$ ; $\continue \leftarrow \FALSE$}
\STATE{\terminate}
	\end{algorithmic}

&

\begin{algorithmic}
\REQUIRE{ $<\!WT,z\!>$ from $X$:}
\STATE{$\nwt\!+\!\!+$ ; $\wt \leftarrow\wt + z$}
\IF{ $\nwt==\nchild$}
\STATE{ \send   $<\!WT,\wt,myId\!>$ via $\pparent$}
\STATE{$\continue \leftarrow \TRUE$}
\ENDIF
\end{algorithmic}

\bigskip

\begin{algorithmic}
{\REQUIRE{ \mbox{$<\!WT2,pwt,Id\!>$} from $X$:}}
\IF{ $\nwt==\nchild$}
\IF{ $pwt\geq \frac{wt}b$}
\STATE{\send $<\!WT',\TRUE\!>$ via $X$}
\STATE{insert $Id$ in $\listH(v)$ ; insert $X$ in $\listP(v)$}
\ELSE
\STATE{\send $<\!WT',\FALSE\!>$ via $X$}
\ENDIF
\ENDIF
\end{algorithmic}

\\
\end{tabular}

\end{algorithm}

We present a distributed convergecast algorithm assuming  the existence of a rooted spanning tree as before with every node having a pointer to its parent. We adapt it to identify heavy and light nodes for $\tz$ preprocessing.
The \emph{weight} $wt$ of a node $v$ is 1 if $v$ is a leaf, or the sum of the weight of its children, otherwise.
 \[wt(v)=\left\{\begin{array}{cl}
 1 & \textrm{ if }v \textrm{ is a leaf}\\
\sum\limits_{u \textrm{ child of }v} \hspace{-0.5cm}wt(u) & \textrm{otherwise} \end{array}\right.\]
For a given constant $b \geq 2$, a node $v$ with parent $p$ is \emph{heavy} if
$wt(v) \geq \frac{wt(p)}{b}$, else $v$ is \emph{light}.

Algorithm~\ref{alg:heavy} computes the weight of the nodes in the tree while also storing
the IDs and ports of its heavy children in its lists $H$ and $P$. It is easy to see that a node can have
at most $b$ heavy children, thus $H$ and $P$ are of size $O(\log n)$.
To compute if the node is a heavy child, it has to wait for its parent to receive the weight of all its children. The parent could then broadcast
or the child continuously sends messages until it receives an answer (message type $WT2$ in Algorithm~\ref{alg:heavy}). Note the broadcast version will accomplish the same task in $O(D)$ rounds with $O(n\Delta)$ messages, so either could be preferable depending on the graph.

\begin{lemma}
Algorithm~\ref{alg:heavy} accomplishes weights computation  in $O(D)$ rounds with $O(nD)$ messages.
\end{lemma}
\nopagebreak
\begin{proof}
 Let sort the node by their depth in the tree, from 0 for the root to $k_0$ for some leaf.
After round $k$, all node of depth $k_0-k+1$ are done.
Then after $k_0+1=O(D)$ rounds, all node are done.

  Each node receives one message from each children and send one to its parent.
  To compute the heavy children there is one more exchange between parent and children.
  This means that there are only a constant number of messages per edge of the tree.
\end{proof}

\subsection{Depth First Walk And Node Relabelling}
\label{sec:dfs}

\renewcommand{\algorithmicrequire}{\textbf{Receive }}

\begin{algorithm}[!htb]
\caption{DFS Traversal Renaming Rules :}
\label{alg:rename}
\footnotesize

\begin{tabular}{p{0.45\linewidth}p{0.45\linewidth}}

\begin{algorithmic}
	\STATE{\init}
		\STATE{$\maxId\leftarrow +\infty$ ; $\fport\leftarrow \bot$}
	\IF{$\Leader$}
		\STATE{$\ivar\leftarrow 0$ ; $\jvar\leftarrow 0$}
		\IF{ $\left|\listH\right| \ge \ivar$ }
			\STATE{ \send $<\!RN, 1\!>$ via $\listP(i)$}
		\ELSE
		\WHILE{\mbox{$\jvar$ not connected}}
	\STATE{$\jvar\leftarrow \jvar+1$} \ENDWHILE
			\STATE{\send $<\!RN, 1\!>$ via port $\jvar$}
		\ENDIF
	\ENDIF
\end{algorithmic}

\medskip

	\begin{algorithmic}
	\REQUIRE{ $<RN, \nextId>$ from $X$:}\STATE\COMMENT{Descending token}
	\IF{$X==\pparent$}
		\STATE{$\ivar\leftarrow 0$ ; $\jvar\leftarrow 0$}
		\IF{$\left|\listP\right|\ge \ivar$}\STATE\COMMENT{Send to an heavy child first}
			\STATE{\send $<\!RN, \nextId\!>$ via $\listP(i)$}
		\ELSE{}\STATE\COMMENT{there is no heavy child}
				\WHILE{{ $\jvar<\Delta$} \AND$\big($\mbox{$\jvar$ not connected} \OR \mbox{$\jvar\in \listP$}  \OR \mbox{$\jvar==\pparent$}$\big)$}
			\STATE{$\jvar\leftarrow \jvar+1$} \ENDWHILE
			\IF{ $\jvar== \Delta$  }
			\STATE\COMMENT{No child, sending back the token}
				\STATE{$\newId \leftarrow nextId$}
				\STATE{$\maxId \leftarrow \newId$}
				\STATE\mbox{\send$<\!RN\!\_UP,\TRUE,\newId, \nextId + 1\!\!>$}
				\STATE{\hspace{0.7cm} via $\pparent$}
			\ELSE{} 
			\STATE{\send $<\!RN, \nextId\!>$ via port $\jvar$}
			\ENDIF
		\ENDIF
	\ELSE
		\STATE\COMMENT{the message didn't come from $\parentBFS$}
		\STATE{ \send $<\!RN\_UP,\FALSE,-1, \nextId \!>$ to $X$}
	\ENDIF
	\end{algorithmic}

&

\begin{algorithmic}
\REQUIRE{ $<\! NEXT\_PORT, p , k \!>$ from $X$:}
	\IF{$X=\pparent$}
	\STATE{$\nport=p$ ; $\listpos=k$}
	\STATE{\terminate}
	\ENDIF
\end{algorithmic}

\bigskip

\begin{algorithmic}
\REQUIRE{$<\!RN\_UP,IsChild,mI,\nextId\!>$~from$X$:}
\STATE\COMMENT{Backtracking token}

\IF{$IsChild$}
	\IF{$\fport==\bot$}
		\STATE{$\fport\leftarrow X$ ; $\maxId \leftarrow mI$}
	\ELSE
		\STATE{\send $<\! NEXT\_PORT, X,\ivar\!>$ via $\last$}
	\ENDIF
\STATE{$\last\leftarrow X$ ; $\ivar\leftarrow \ivar +1$}
\ENDIF
\medskip
\IF{ $\left|\listH\right|\ge \ivar$}\STATE\COMMENT{Send to the next heavy child first}
	\STATE{ \send $<\!RN, \nextId\!>$ via $\listP(i)$}
\ELSE{}\STATE\COMMENT{there is no more heavy child}
				\WHILE{{ $\jvar<\Delta$} \AND$\big($\mbox{$\jvar$ not connected} \OR \mbox{$\jvar\in \listP$} \OR \mbox{$\jvar==\pparent$}  $\big)$}
		\STATE{ $\jvar\leftarrow \jvar+1$} \ENDWHILE
	\IF{$\jvar== \Delta$}
		\STATE{ \send $<\! NEXT\_PORT, \bot, \ivar-1\!>$ via $X$}
				\STATE{$\newId \leftarrow \nextId$}
		\IF{$\Leader$}
			\STATE{\terminate}
		\ELSE\STATE\COMMENT{No more child, sending back the token}
				\STATE{\mbox{\send$<\!RN\!\_UP,\TRUE,\maxId, \nextId+1\!>$}
				\STATE{\hspace{0.7cm} via $\pparent$}}
				\ENDIF
	\ELSE \STATE{\send $<\!RN, \nextId \!>$ via port $\jvar$}
	\ENDIF
\ENDIF
\end{algorithmic}
\\

\end{tabular}
\end{algorithm}

The next step in the preprocessing of $\tz$ is to relabel the nodes using the spanning tree
computed in the previous section. The labels are computed through a post-order DFS walk on the tree,
prioritizing the walk towards heavy children.
In the algorithm,
the root starts the computation,
sending the token with the ID set to 1 to its first heavy child.
Once a node gets back the token from all its children,
it takes the token's ID as its own, increments the token's ID and sends to its parent.
Note that in our algorithm, each node $v$ has to try all its ports when passing the token (except the port connected
to its parent) since $v$ cannot `remember' which ports connect to the spanning tree.
Our solution to this problem is to ``distribute'' that information among the
children. This problem is solved while performing the DFS walk.
Each node $v$, being the $l$-th child of its parent $p$,
has a local variable $\nport$, which stores the port number of $p$ connecting it with its $(l+1)$-th child.
This \emph{compact} representation of the tree will allow us to be round optimal in the next section.

{\begin{lemma}
Algorithm \ref{alg:rename} relabels the nodes in a pre-order DFS walk in $O(m)=O(n^2)$ rounds,
using $O(m)=O(n^2)$ messages overall.
\end{lemma}
\nopagebreak
\begin{proof}
The token walks over the graph by exploring all the edges (not just the tree edges),
each edge are used at most 4 times.
During each round, exactly one node sends and receives exactly one message.
\end{proof}
}

\subsection{Computing Routing Labels}
\label{subsec:prepoc}


\renewcommand{\algorithmicrequire}{\textbf{Receive }}

\begin{algorithm}[htb]
\caption{Computing Routing Labels with $O(\log^2 n)$ sized messages:}
\label{alg:Lpath2}
\footnotesize

\begin{tabular}{p{0.45\linewidth}p{0.45\linewidth}}

	\begin{algorithmic}
	\REQUIRE{$<\! RL, P, Y \!>$ from port $X$:}
	\IF{$\pparent==X$}
		\IF{$\heavy$}
			\STATE{$\lightPath \leftarrow P$}
		\ELSE
			\STATE{$\lightPath \leftarrow P \addList Y$}
		\ENDIF
		\STATE{$\routingLabel \leftarrow (\newId, \lightPath)$}
		\FORALL{port $Z$}
			\STATE{\send $<\! RL, \lightPath, Z \!>$ via $Z$}
	\ENDFOR
	\STATE{\terminate}
	\ENDIF
	\end{algorithmic}
	&
\begin{algorithmic}
	\STATE{\init}
	\IF{$\Leader$}
	\STATE{$\lightPath \leftarrow \emptyList$}
	\STATE{$\routingLabel \leftarrow (\newId, \lightPath)$}
	\FORALL{port $X$}
		\STATE{\send $<\! RL, \lightPath, X \!>$ via $X$}
	\ENDFOR
	\ENDIF
\end{algorithmic}
\\
\end{tabular}
\end{algorithm}

We now have enough information (a leader, a BFS spanning tree, node weights, DFS labels)
to produce routing labels in $\tz$, and hence, to complete the preprocesing.
For a node $v$, its \emph{light path} is the sequence of port numbers for light nodes in the path from the root to $v$.
The routing label of $v$ in $\tz$ is the pair $(\newId, \lightPath)$,
where $\newId$ is its DFS label and $\lightPath$ its light path.
The second routing table entry for the root is empty.

A simple variant of Algorithm~\ref{alg:tree} computes the routing labels if $O(\log^2 n)$ sized messages are permitted (Algorithm~\ref{alg:Lpath2}),
otherwise a slower variant can do the same with $O(\log n)$ messages (Algorithm~\ref{alg:Lpath1}).
For the $O(\log^2 n)$ size variant, the root begins by sending its path(\emph{empty}) to each port  $X$ along with the port number $X$.
When a node receives a message $<RL, path, X>$ from its parent,
it sets its light path to $path\addList X$, if it is light, otherwise to $path$ only, producing its routing label.
Then, for each port $X$, it sends its light path together with the port number $X$.
For the $O(\log n)$ size variant (Algorithm~\ref{alg:Lpath1}), every light node receives from its parent the port number it is on (say, port $X$) and then does a broadcast labeled with $X$. The root also broadcasts a special message. Ever receiving node appends a received $X$ to its $path$ incrementing its light level and terminating when receiving the root's message.

\renewcommand{\algorithmicrequire}{\textbf{Receive }}

\begin{algorithm}[htb]
\caption{Computing Routing Labels with $O(\log n)$ sized messages:}
\label{alg:Lpath1}
\footnotesize

\begin{tabular}{p{0.45\linewidth}p{0.45\linewidth}}

			\begin{algorithmic}
	\STATE{\init}
		\STATE{$\lpath \leftarrow\emptyList$}
	\FORALL{port $X$}
	\STATE{\send $<\!PORT,X,\Leader\!>$ via $X$}
	\ENDFOR
		\IF{  $\Leader$}
			\STATE{\terminate}
		\ENDIF
	\end{algorithmic}

\medskip

	\begin{algorithmic}
	\REQUIRE{$<\!PORT,Y,isL\!>$ from port $X$:}
		\IF{$\pparent==X$ \AND \NOT $\heavy$}
			\STATE{$\lpath \leftarrow Y\addList\lpath$}
			\IF{$\nchild>0$}
				\STATE{\bdcast $<\!PORT2,Y,isL\!>$except\,$\pparent$}
			\ENDIF
		\ELSIF{$isL$}
				\STATE{\bdcast $<\!PORT2,\bot,isL\!>$except\,$\pparent$}
		\ENDIF
		\IF{  $isL$}
			\STATE{\terminate}
		\ENDIF
	\end{algorithmic}

		&

	\begin{algorithmic}
	\REQUIRE{$<\!PORT2,Y,isL\!>$ from port $X$:}
		\IF{$\pparent==X$}
			\IF{$Y\neq \bot$}
			\STATE{$\lpath \leftarrow Y\addList\lpath$}
			\ENDIF
			\IF{$\nchild>0$}
				\STATE{\bdcast $<\!PORT,Y,isL\!>$except\,$\pparent$}
			\ENDIF
		\ENDIF
		\IF{ $isL$}
			\STATE{\terminate}
		\ENDIF
	\end{algorithmic}

\\
\end{tabular}
\end{algorithm}

\begin{lemma}
Algorithm~\ref{alg:Lpath2} computes the routing labels of $\tz$ in $O(D)$ rounds using $O(m)$ messages of $O(\log^2 n)$ size.
\end{lemma}
\nopagebreak
\begin{proof}
Each node receive at most one message from each neighbors (and take into consideration only the one coming from its parent). And the longest a message can travel is from the root to a leaf. Each node sends only 1 message to the server.
This requires then $O(m)$ messages and $O(D)$ rounds.
\end{proof}

\begin{lemma}
Algorithm~\ref{alg:Lpath1} computes the routing labels of $\tz$ in $O(D)$ rounds using $O(mD)$ messages of $O(\log n)$ size.
\end{lemma}\nopagebreak[4]\begin{proof}
During the initialization round, each node send trhough each of its port the corresponding port number.
during the first round each node receive those port number and consider only the one coming from its parent.
If it is a light child, it broadcasts this number in its subtree.
This way each node will receive at different time but in order all the chain of port to a light child from its parent to the root.
The messages initiated by the root are special and forwarded even by its heavy children since it is a termination signal.

At each round at most one message is send on any edge.
And the longest travel of a message is from the root to the furthest leaf.
Therefore this requires then $O(mD)$ messages and $O(D)$ rounds.
\end{proof}


\section{Compact Forgiving Tree}
\label{sec:cftree}

%

 Section~\ref{sec: cftz overview} gives an overview of $\cft$.  As it stated, the central idea is a node's $\will$ (its $\RT$) which needs to be pre-computed before an adversarial attack. \cite{CDT2018-CompactFTZ}  has only a distributed non-compact memory preprocessing stage\footnote{Once the non-compact memory preprocessing stage is completed,
each process uses only compact memory during the execution of the algorithm.} in which, in a single round of communication,
each node gathers all IDs from its children, locally produces its $\will$, and then, to each child, sends a part of its $\will$, called $\willportion$ or \emph{subwill}, of size $O(\log n)$. Computing the $\will$ with compact memory is a challenging problem
as a node might have $\Omega(n)$ neighbours making its $\will$
 of size $\Omega(n \log n)$. Thus, to compute this information in a compact manner,
we need a different approach, possibly costing more communication rounds.
Remarkably, as we show below,
one round is enough to accomplish the same task in the $\CModelAbbrev$ model, with deterministic reads.
The solution is made of two components: a \emph{local compact function} in Section~\ref{subsection-computing-labels}
that efficiently computes parts of labelled half-full trees of size $O(n \log n)$ using only $O(\log n)$ memory,
and a \emph{compact protocol} in Section~\ref{subsection-computing-will} that solves the distributed problem of
transmitting the $\willportions$ to its children in a single round.

\subsection{Computing Half-Full Trees with Low Memory }
\label{subsection-computing-labels}

Half-full trees~\cite{FG-DCJournal2012} (which subsume balanced binary trees), redefined below, are the basis for computing the $\will$ of a node in CompactFT.
At the core of the construction is a labelling of the nodes with good properties that allows
to have $\willportions$ of size $O(\log n)$ to the children of a node.
Roughly speaking, a half-full tree is made of several full binary trees, each with a
binary search tree labelling in its internal nodes.
In what follows we show how to compute properties of that label

\subsubsection{Computing labels of full binary trees}
Given a power of two, $2^x$, consider the full binary tree with $2^x$ leaves defined recursively as follows.
The root of the tree is the string $0$, and each node $v$ has left child $v \, 0$ and right child $v \, 1$.
It is easy to see that the nodes at height $h$ are the binary representation of $0, \hdots, 2^{x-h}-1$.
We write $\tilde{v}$ the integer represented by the chain $v$.
Moreover, for any node $v$, its left and right children represent the number $2\tilde v$ and $2\tilde v+1$, respectively.
Let $B(2^x)$ denote the previous tree.
We now define a function $\ell$ used in CompactFT that labels the nodes of $B(2^x)$ in the space $\interval 0 {2^x-1}$.
Of course the labelling is not proper but it has nice properties that will allow us to compute it
using low memory.

Consider a node $v$ of $B(2^x)$.
Let $h_v$ denote the height of $v$ in $B(2^x)$. Then, we define $\ell$ as follows:~if  $h_v = 0$, $\ell(v) = \tilde v$, otherwise $\ell(v)=2^{h_v-1}-1 + \tilde v \, 2^{h_v}$.

\begin{figure}[tb]
\centerline{\includegraphics[width=15cm]{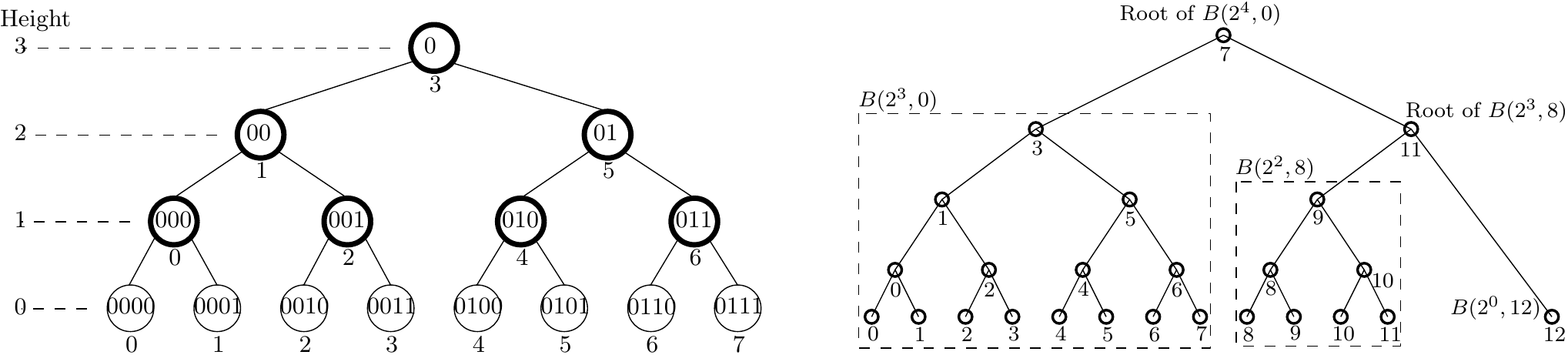}}
\caption{\footnotesize The tree at the left is $B(2^3)$ with its labeling $\ell$. Each circle shows in its interior the binary identifying the vertex and its decimal value.
Near each node appears its label $\ell$. Non-leaf nodes correspond to bold line circles. The tree at the right is the half-full tree $HT([0,12])$ with its labeling $\ell'$.
\label{figure-trees}
}
\end{figure}

In words, if $v$ is of height $0$, its label is simply $\tilde v$, otherwise its label is computed using
a base number, $2^{h_v-1}-1$, plus $\tilde v$ times an offset, $2^{h_v}$.
{Figure~\ref{figure-trees} (left)} depicts the tree $B(2^3)$ and its labelling $\ell$.
Note that the internal nodes have a binary search tree labelling.

\begin{lemma}
\label{lemma-prop-labelling}
Let $B(2^x)$ be a non trivial tree --with $x>0$.
For every vertex $v$, $\ell(v) \in \interval 0 {2^x-1}$.
 For the root $r$, $\ell(r) = 2^{x-1}-1$.
Consider any $y \in \interval 0 {2^x-1}$.
There is a unique leaf $v$ of $B(2^x)$ such that $\ell(v) = y$.
 If $y \leq 2^x-2$, there is a unique non-leaf $u$ of $B(2^x)$ such that $\ell(u) = y$,
 and there is no non-leaf $u$ of $B(2^x)$ such that $\ell(u) = 2^x-1$.
\end{lemma}

\begin{proof}
Let $v$ be a node of $B(2^x)$.
As explained above, $\tilde v \in \interval 0 {2^{x-h_v}-1}$.
It is clear that $\ell(v) \geq 0$.
If $h_v = 0$, then $\ell(v) = \tilde v \leq 2^x-1$.
Else $\ell(v) = 2^{h_v-1}\!-\!1\!+\!\tilde v \, 2^{h_v}
			\!\leq
			2^x\!-\!1\!-\!2^{h_v-1}
< 2^x\!-\!1$.

The root $r$ of $B(2^x)$ has height $h_r=x$ and $\tilde r= 0$,
hence, by definition, $\ell(r) = 2^{x-1}-1$.
Now, consider any $y \in\interval 0 {2^x-1}$.
Since all leaves are at height $0$, there is a unique leaf $v$ with $\ell(v) = y$.
Suppose that $y \leq 2^x - 2$.
There exists a unique interger factorization of $y+1$ then there exists unique $h\ge 0$ and $p\ge 0$ such that $y+1=2^{h}(2p+1)$.
This decomposition can be easily obtained from the binary representation of $y+1$.
By construction, we have $h\le \log(y+1) < \log(2^x)=x$ then $h+1\leq x$ and we have $2p < 2^{x-h}$ then $p\leq 2^{x-h-1}-1$.
Let consider the unique (non-leaf) node $u$ such that $\tilde u=p$ and $h_u=h+1\ge 1$.
It means that $u$ is the unique node such that $\ell(u)=y$.
Finally, there is no non-leaf $u$ of $B(2^x)$ such that $\ell(u) = 2^x-1$ because
we just proved that each element of $\interval 0 {2^x-2}$  has a unique inverse image under $\ell$.
Since the number of non-leaf node is exactly $2^x-1=\left|\interval 0 {2^x-2} \right|$,
there is no non leaf node $u$ such that $\ell(u)=2^x-1$.
\end{proof}

By Lemma~\ref{lemma-prop-labelling}, when considering the labelling $\ell$,
each $y \in \interval 0 {2^x-1}$ appears one or two times in $B(2^x)$,
on one leaf node and at most on one non-leaf node. Thus, we can use the labelling $\ell$ to
unambiguously refer to the nodes of $B(2^x)$.
Namely, we refer to the leaf $v$ of $B(2^x)$ with label $\ell(v) = y$ as
\emph{leaf} $y$,
and, similarly, if $y \leq 2^x-2$, we refer to the non-leaf $u$ of $B(2^x)$ with label
$\ell(u) = y$ as \emph{non-leaf} $y$.
By abuse of notation, in what follows $B(2^x)$ denotes the tree itself and its labelling $\ell$ as defined above.
The following lemma directly follows from the definition of $\ell$.

\begin{lemma}
\label{lemma-parent-children}
Let $B(2^x)$ be a non trivial tree ($x\ge 1$).
Consider any $y \in \interval 0 {2^x-1}$.
The parent of the leaf  $y$ is the non-leaf $2 \lfloor \frac{y}{2} \rfloor$.
If $y \leq 2^x-2$, then let $y = 2^i - 1 + z \, 2^{i+1}$.
If $i\le x-2$, the parent of the non-leaf  $y$ is the non-leaf $2^{i+1}-1\! +\! \lfloor \frac{z}{2} \rfloor \, 2^{i+2}$.
If $i\ge 1$, the left and right children of the non-leaf  $y$ are the non-leafs $2^{i-1}\!-\! 1\! + 2z \, 2^i$ and $2^{i-1}\!-\! 1\! + (2z+1) \, 2^i$, respectively.
If $i=0$, the left and right children of the non-leaf $y$ are the leafs $y$ and  $y+1$.
\end{lemma}

The proof of Lemma~\ref{lemma-prop-labelling}
 shows how to quickly represent a non-leaf node $v$ with
$\tilde v  \in\interval 0 {2^x-2}$ in its form $\tilde v = 2^i-1 + \lfloor \frac{\tilde u}{2} \rfloor \, 2^{i+1}$ so that
one can easily compute its parent and children, using Lemma~\ref{lemma-parent-children}.
Given a value $y  \in \interval 0 {2^x-1}$,
function {\SearchBT} in Algorithm~\ref{alg-search-bt} returns the parent of the leaf node
$v$ of $B(2^x)$ and the parent and children of the non-leaf node $v'$ of $B(2^x)$ such that $y=\tilde v=\tilde{v'}$.
The correctness of the function directly follows from Lemma~\ref{lemma-parent-children}.
Note that the function uses $o(x)$ memory: $O(\log x)$ bits to represent $y$ and
a constant number of variables with $O(\log x)$ bits.

Let $B(2^x, a)$ denote the tree $B(2^x)$ together with the labelling
$\ell'(v) = \ell(v) + a$. Clearly, $\ell'$ labels the nodes of $B(2^x, a)$ in the space $\interval a {a+2^x-1}$.
For ease of representation, we use $B(2^x)$ to represent $B(2^x,0)$ (i.e. $B(2^x)$ with labelling $\ell(v)$) in the discussion that follows.

{
\begin{remark}
\label{remark-binary-trees}
The left subtree of the root of $B(2^x)$ is $B(2^{x-1})$ and its right subtree is $B(2^{x-1}, 2^{x-1})$.
Thus, the left subtree of the root of $B(2^x, a)$ is $B(2^{x-1}, a)$ and
its right subtree is $B(2^{x-1}, 2^{x-1}+a)$.
\end{remark}
}

\subsubsection{Computing labels of half-full trees:}
Here, we give a compact function to return the requisite labels from a half-full tree.

\emph{Half-full trees} are the basis for constructing the will portions of the neighbours of a node
in CompactFT. Here we want to compute that data using low memory.

\begin{definition}\cite{FG-DCJournal2012}
\label{definition-hf-tree}
Consider an integer interval $S = \interval a {b}$. The \emph{half-full tree with leaves in} $S$,
denoted $HT(S)$, is defined recursively as follows.
If $|S|$ is a power of two then $HT(S)$ is $B(|S|, a)$.
Otherwise, let $2^x$ be the largest power of two smaller than $|S|$.
Then, $HT(S)$ is the tree obtained by replacing the right subtree of the root of $B(2^{x+1}, a)$
with $HT(\interval{a+2^x}{b})$.
The nodes of $HT(S)$ have the induced labelling $\ell'$ of every $B(*, *)$ recursively used for defining the half-full tree.
\end{definition}.

 {Figure~\ref{figure-trees} (right)} depicts the half-full tree $HT(\interval0{12})$ and its induced $\ell'$ labelling.
The following lemma states some properties of the $\ell'$ labelling of a half-full tree.

 \begin{lemma}
 \label{lemma-props-labels}
 Consider a half-full tree $HT(\interval a{b})$.
 For every node $v$ of $HT(\interval a{b})$, $\ell'(v) \in \interval a{b}$.
 For the root $r$ of $HT(\interval a{b})$, $\ell'(r) = 2^x-1+a$,
 where $2^x$ is the largest power of two smaller than $b-a+1$.
 For every $y \in [a, b]$, there is a unique leaf $v$ of $HT(\interval a{b})$ such that $\ell'(v) = y$,
 and if $y \in \interval a{b-1}$, there is a unique non-leaf $v$ of $HT(\interval a{b})$ such that $\ell'(v) = y$.
 \end{lemma}

 {
 \begin{proof}
 First, it directly follows from Definition~\ref{definition-hf-tree} that for every $v$ of $HT(\interval a{b})$,
 $\ell'(v) \in\interval a{b}$.
 By Lemma~\ref{lemma-prop-labelling}, $2^x-1$ is the root of $B(2^{x+1})$,
 and thus $2^x-1+a$ is the root of $B(2^{x+1}, a)$, and consequently the root of $HT(\interval a{b})$.

 By construction and using Remark~\ref{remark-binary-trees}, if $y\in\interval a{a+2^x-1}$, there is a unique leaf $v$ in the right subtree of the root such that $y=\ell(v)$ and for all leaf $v'$ outside this the subtree we have $\ell(v')\in\interval{a+2^x}b$. By induction on the HF construction, if $y\in\interval {a+2^x+1}b$, there is a unique leaf $v$ in the left subtree of the root such that $y=\ell(v)$ and for all leaf $v'$ outside this the subtree we have $\ell(v')\in\interval a{a+2^x}$.

 By the inductive construction of the $HT(\interval a{b})$, all leaf nodes belongs to one full binary subtree, and all the full binary subtrees cover disjoint intervals.
 By  Remark~\ref{remark-binary-trees}, there exists a unique leaf node $v$ such that $\ell(v)=y$.

 \end{proof}
 }

 \begin{figure}[tb]
 \centerline{\includegraphics[width=10.5cm]{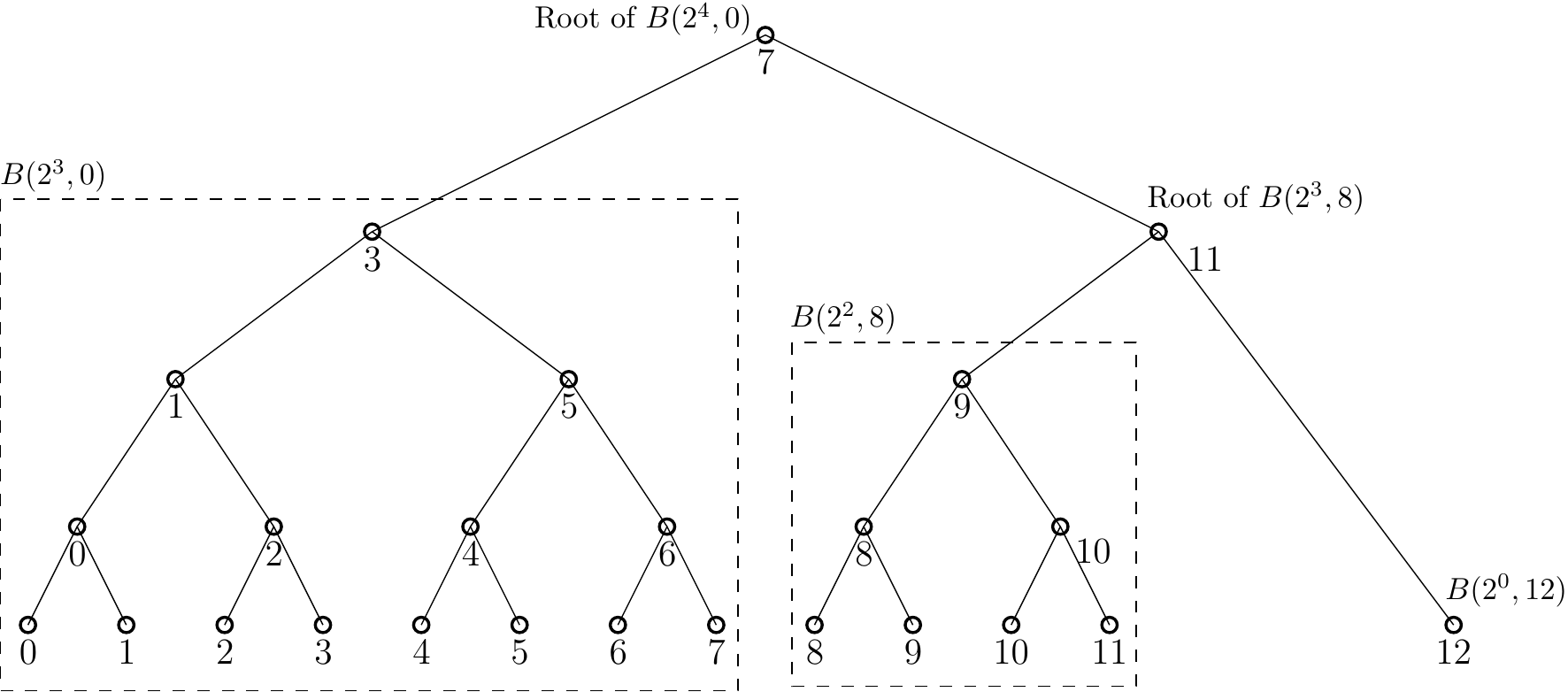}}
 \caption{The half-full tree $HT([0,12])$ with its labeling $\ell'$.}
 \label{figure-half-full-tree}
 \end{figure}

As with full binary trees, by Lemma~\ref{lemma-props-labels}, when considering the labelling $\ell'$ of $HT(\interval a{b})$,
each $y \in \interval a b$ appears one or two times in $HT(\interval a{b})$,
on one leaf node and at most on one non-leaf node. Therefore, those two nodes in the half-full tree can be
unambiguously referred as \emph{leaf node} $y$ and \emph{non-leaf node} $y$.
The very definition of half-full trees,~Definition~\ref{definition-hf-tree}, and Lemma~\ref{lemma-props-labels}
suggest a natural low memory (sublinear on the size of the interval) recursive function that obtains the parent and children of a node
in $HT(\interval a{b})$. Such a function, \SearchHF,
appears in Algorithm~\ref{alg-search-hf}.

\begin{lemma}
\label{lm: searchhf}
Function {\SearchHF}($y, a, b$) computes the parent and children of leaf and non-leaf \mbox{$y \in \interval a{b-1}$} in~$HT(\interval a{b})$
using $O(\log b)$ space.
\end{lemma}

\subsection{Computing and distributing will portions in one round}
\label{subsection-computing-will}

Among other things, in Section~\ref{sec: basictz}, we have computed a spanning tree $T$ of the original graph $G$.
Here we present a one-round compact protocol that, for any node $x$, computes and sends to each child of $x$ in $T$
its corresponding will portion.
Let $\delta$ denote the number of children of $x$ in $T$.
The will of $x$ is the half-full tree $HT([0,\delta-1])$, where each label $l$ is replaced with the ID
of the $l$-th child of $x$ in $T$. Let $RT(x)$ denote this tree with the IDs at its nodes.
Thus each child of $x$ with ID $y$ appears two times in $y$, one as leaf node and one as a non-leaf node,
and the subwill of $y$ in $RT(x)$ is made of the parent of the leaf $y$ and the parent and children of the non-leaf $y$ in $RT(x)$.
This is the information that $x$ has to compute and send to $y$. We can efficiently compute the subwill of a child using a
slight adaptation of the function {\SearchHF} defined in the previous subsection.

The representation of $T$ is compact: $x$ only knows its number of children in $T$ and the port of its first children
(the ports of its children do not have to be contiguous). Additionally, the $l$-th child of $x$ has the port number of $x$, $nxt\_port$,
that is connected $(l+1)$-th child of $x$.
In our solution, shown in Algorithm~\ref{alg:will}, $x$ first indicates to all its children to send its ID and $nxt\_port$ so that this data  is in the in-buffers of $x$.
Then, with the help of the $nxt\_port$, $x$ can sequentially read a collect the IDs of its children,
and in between compute and send will portions.
In order to be compact, $x$ has to ``forget'' the ID of a child as soon as it is not needed anymore for computing the will portion
of a child (possibly the same or a different one).
For example, if $\delta = 13$, then $x$ uses the half-full tree $HT([0,12])$ in Figure~\ref{figure-trees},
and the label $l$ in $HT([0,12])$ denotes to the $l$-th child of $x$ in $T$.
After reading and storing the IDs of its first four children (corresponding to $0, 1, 2, 3$),
$x$ can compute and send the subwill of its first and second children (0 and 1).
The leaf 0 in $HT([0,12])$ has parent non-leaf 0 while the non-leaf 0 has parent non-leaf 1 and children leaf 0 and leaf 1.
Similarly, the leaf 1 has parent non-leaf 0 while the non-leaf 1 has parent non-leaf 3 and children non-leaf 0 and non-leaf 2.
Moreover, at that point $x$ does not need to store anymore the ID of its first child because (leaf or non-leaf) 0 is not part of
any other will portion. An invariant of our algorithm is that, at any time, $x$ stores at most four IDs of its children.

The rules appear in Algorithm~\ref{alg:will}.
The algorithm uses function {\bf SubWill}, which computed the subwill of a child node using function the
compact function \SearchHF.

{
\begin{algorithm}
\caption{Calculates the parent and children of leaf and non-leaf $y \in \interval 0  {2^x-1}$ in~$B(2^x)$.}
\label{alg-search-bt}
\footnotesize
\vspace{-1em}
\begin{flushleft}
{\bf Function \SearchBT($y, 2^x$)}
\end{flushleft}
\vspace{-1.5em}
\begin{algorithmic}

\IF{$x=1$}
	\RETURN  $\langle 0, \bot , 0 , 1 \rangle$
\ELSE
	\IF {$y=2^{x-1}-1$}
	\STATE $\langle P, P', L', R' \rangle\leftarrow  \langle 2^{x-1}-2 , \bot , 2^{x-1}-1-2^{x-2} , 2^{x-1}-1+2^{x-2}\rangle$ \COMMENT $y$ is the root
\ELSIF{$y<2^{x-1}-1$}
			\STATE $\langle P, P', L', R' \rangle\leftarrow $ {\SearchBT}($y, 2^{x-1}$) \COMMENT $y$ is in the left subtree
	\ELSE
			\STATE $\langle P, P', L', R' \rangle\leftarrow $ {\SearchBTDelta}($y,2^{x-1}, 2^{x-1}$)  \COMMENT $y$ is in the right subtree
	\ENDIF
\ENDIF
	\IF {$P'=\bot$}
		\STATE  $P' \leftarrow  2^{x-1}-1$ \COMMENT $y$ is in the root of one of the two subtrees
	\ENDIF
	\RETURN $\langle P, P', L', R' \rangle$
\end{algorithmic}

\medskip

	{\bf Function \SearchBTDelta($y, 2^x, a$)}
\begin{algorithmic}
	\STATE $\langle P, P', L', R' \rangle \leftarrow \, ${\SearchBT}($y-a, 2^x$)
	\RETURN{$\langle P+a, P'+a, L'+a, R'+a \rangle$ }
\end{algorithmic}
\end{algorithm}

\begin{algorithm}
\caption{Calculates the parent and children of leaf and non-leaf $y \in \interval a{b-1}$ in~$HF(\interval a{b})$.}
\label{alg-search-hf}
\footnotesize
\vspace{-1em}
\begin{flushleft}
{\bf Function \SearchHF($y, a, b$)}
\end{flushleft}
\vspace{-1.5em}
\begin{algorithmic}
	\IF{$b-a=2^x$ for some $x$}
		\RETURN{{\SearchBTDelta}($y, 2^x, a$)}  \COMMENT The HT is actually a BT
	\ELSE
	\STATE $\langle P, P', L', R' \rangle\leftarrow  \bot , \bot , \bot , \bot \rangle$
\STATE  $x = \left\lfloor \log_2(b-a)\right\rfloor$
						\COMMENT {let} $2^x$ be the largest power of two smaller than $b-a+1$
\STATE  $z = \left\lfloor \log_2(b-a-2^x)\right\rfloor$
				\COMMENT {let} $2^z$ be the largest power of two smaller than $b-a-2^x$
\IF {$y=a+2^{x}-1$}
	\STATE $\langle P, P', L', R' \rangle\leftarrow  \langle  2^x-2 , \bot , 2^{x}-1-2^{x-1}, 2^{x}-1+2^{z}\rangle$ \COMMENT $y$ is the root
\ELSIF{$y<a+2^{x-1}-1$}
			\STATE $\langle P, P', L', R' \rangle\leftarrow$ {\SearchBTDelta}($y, 2^{x-1}, a$) \COMMENT $y$ is in the left subtree
	\ELSE
			\STATE $\langle P, P', L', R' \rangle\leftarrow$ {\SearchHF}($y,a+2^x, b$)  \COMMENT $y$ is in the right subtree
	\ENDIF
\ENDIF
		\IF {$P'=\bot$}
			\STATE  $P' \leftarrow  2^{x-1}-1$ \COMMENT $y$ is in the root of one of the two subtrees
		\ENDIF
	\RETURN $\langle P, P', L', R' \rangle$
\end{algorithmic}

\end{algorithm}
}

\begin{algorithm}
\caption{Calculates the subwill associate to the child $y \in \interval 0{b-1}$.}
\label{fct-subwill}
\footnotesize
\vspace{-1em}
\begin{flushleft}
{\bf Function SubWill($y, b, parent$)}
\end{flushleft}
\vspace{-1.5em}
\begin{algorithmic}
	\STATE{$\langle P, P', L', R' \rangle \leftarrow$  {\SearchHF($y, 0, b$)}  }
	\STATE  $x = \left\lfloor \log_2(b)\right\rfloor$ \COMMENT{$2^x$ is the largest power of two smaller than $b$}
	\IF{$y=b-1$}
	\STATE{}\COMMENT {$y$ is in the root of one of the two subtrees}
			\STATE{$ P' \leftarrow \, parent$ ; $  L' \leftarrow \, 2^x-1 $ }
	\ELSIF{$y=2^x-1$} \STATE{}\COMMENT{$y$ is the root}
			\STATE{$ P' \leftarrow \,b-1 $ }
	\ENDIF
	\RETURN{$P, P', L', R'$ }
\end{algorithmic}
\end{algorithm}

\renewcommand{\algorithmicrequire}{\textbf{Receive }}

\begin{algorithm}[!htb]
\caption{Wills}
\label{alg:will}
\footnotesize

\begin{tabular}{p{0.45\linewidth}p{0.45\linewidth}}

\begin{minipage}{\linewidth}
\begin{algorithmic}
	\STATE{\init}
	\IF{DFS walk is over :}
		\STATE{$\current \leftarrow \fport$ ; $\vark \leftarrow 0$}
		\IF{not $\Leader$}
		\STATE{ \send  $<\!MYId, myId, \nport, \listpos \!>$ via $\pparent$}
		\ENDIF
	\ENDIF
\end{algorithmic}

\bigskip

\begin{algorithmic}
\REQUIRE{$<\!WILL,p,p_h,c_l,c_r,bool\!>$ from $X$}
\STATE{$\nparent \leftarrow p$ ; $\nparenth \leftarrow ph$}
\STATE{$\nlchild \leftarrow c_l$ ; $\nrchild \leftarrow c_r$}
\STATE{$\heirRT \leftarrow bool$}
\STATE{\terminate}
\end{algorithmic}
\end{minipage}

&
\begin{minipage}{\linewidth}
\begin{algorithmic}
\REQUIRE{$\!\!<\!MYId,z,\ncurrent,\_\!>$from\,$\current$:}
	\STATE{$\listNode[\vark]\leftarrow [0,z,\current]$}
	\STATE{$\listWill[\vark]\leftarrow {\bf SubWill}(\vark,\nchild,\parentBFS)$}
	\FORALL{$\jvar \in \listWill[\vark]\cup\{\vark\}$}
	\IF{$\max\left(\listWill[\jvar]\right)\le \vark$}
	\STATE{$p,p_h,c_l,c_r=\listWill[\jvar]$}
	\STATE{\send  $<\!WILL, \listNode[p][1],$$\listNode[p_h][1],$$\listNode[c_l][1],$ $\listNode[c_r][1],$${[\nchild\!-\!1\!==\!\vark]}\!>$via\,$\listNode[\jvar][2]$ }
	\STATE{$free(\listWill[\jvar])$}
	\FORALL{$x\in\{p,p_h,c_l,c_l\}$}
	\STATE{$\listNode[x][0]+\!+$}
	\IF{$\listNode[x][0]==4$}\STATE{$free(\listNode[x])$}\ENDIF
	\ENDFOR
	\ENDIF
	\ENDFOR
	\STATE{$\current \leftarrow \ncurrent$ ; $\vark+\!+$}
	\end{algorithmic}
\end{minipage}

\end{tabular}

\end{algorithm}

\pagebreak
\begin{lemma}
All the subwills are correctly computed and sent in 1 round.
\end{lemma}

\begin{proof}
By construction, all the children form a chain starting with $\fport$
and going on via the pointers $\nport$. If those are correct, all the in-buffers
corresponding to children are read during this one round.
Thus $\vark$ goes from $0$ to $\delta-1$ and  $\current$ from $\fport$ to the last one through every port.
This means that for every $\vark$, at some point $Node[\vark]$ and $Will[\vark]$ are filled.

For each $\vark$, $Will[\vark]$ is freed only after use and $Node[\vark]$ only after 4 uses.
No $Node[\vark]$ need more than 4 uses.

Let $0\le k < \delta$.
Let $k_0\le k_1\le k_2\le k_3$ the nodes used in $Will[\vark]$.
Eventually $Node[k_3]$ will be filled. At this point $Node[k_i]$ is filled for $i=0,1,2,3$
and none of them were freed, since there were still at least one missing use.
The subwill can be correctly sent to the $\vark^{th}$ child.

In conclusion, each child's port is read and each subwill is correctly computed and sent in one round.
\end{proof}

\begin{lemma}
 At any time, the memory contains at most $5 \log \delta$ IDs (node or port) and subwills.
 \end{lemma}

\begin{proof}
We define the uncompleted edges as the edges $(u,v)$
such that $u$ is already registered in $Node$ and $v$ is not.
For each of those edges we have to remember informations about $u$
(its id, its port id, its subwill with the ids of corresponding nodes (except $v$, then at most 3 nodes id)).
The size of this information is at most $5$ Ids of size $O(\log n)$.

If $u$ is linked to the $k_0^{th}$ port and $v$ to the $k_1^{th}$ port, then we have $k_0\le k < k_1$.
By construction, $k_1-k_0=2^h-\sum_{l=h'}^{h-1} 2^l=2^{h'}$ for some $0\le h'\le h\le \log \delta$.
For any value of $h$, there can be only one such edge
(this edge being between the port number $2^{h+1}\lfloor\frac {k-2^{h-1}}{2^{h+1}}\rfloor- 2^{h-1}-1$
and a port whose number is smaller or equal to $2^{h+1}\lfloor\frac {k-2^{h-1}}{2^{h+1}}\rfloor+ 3.2^{h-1}-1$).
Thus there are at most $\log n$ such edges.

At any moment the total amount of used memory is $O(\log^2 n)$.
\end{proof}

\subsection{Computing and distributing will parts with adverserial reads}
\label{subsection-computing-will-adv}
The previous protocol works only in the deterministic reads case.
However, it can be adapted to the adversarial reads case at the cost of some more rounds.
Instead of computing all subwills in one round, we now compute one subwill per round.
At round $k$, a node computes the subwill of its $k$-th child.
To do so, it reads all the ports and stores only the needed IDs for computing the subwill of its $k$-th child,
and then it sends it to the child.
%


\section{Related Work}

In the \emph{network finite state machine} model~\cite{EmekW13}, weak computational devices with constant local memory communicate in an asynchronous network.
Any node only broadcasts symbols from a constant size alphabet and each time it reads its ports (all of them) can only distinguish up to a constant number of occurrences. They show probabilisitic solutions to MST and 3-coloring of trees in this model.


In the \emph{beeping} model of communication~\cite{DegesysRPN07}, nodes execute synchronous rounds, and in any round, each process decides to ``beep'' and send but not receive
or to stay silent and listen. A node that listens obtains a single bit encoding if at least one of its neighbours beeped in that round.~\cite{GilbertN15} have shown that there are probabilistic solutions to the leader election problem in the beeping
model for the complete graph topology, in which each node is a state machine with constant states.
The aforementioned solutions in~\cite{GilbertN15} imply compact probabilistic solution in our $\CModelAbbrev$ model.

As far as we know, the computational power of the CONGEST models ($O(poly \log n)$ sized messages)~\cite{peleg} has never been studied when the local memory
of the nodes is restricted too. However,~\cite{DruckerKO13}  has studied the difference between nodes performing only broadcasts or doing unicast, showing that the unicast model is strictly more powerful.
~\cite{BeckerARR15} studied the general case where nodes are restricted to sending some number of distinct messages in a round. It'll be interesting to see where $\CModelAbbrev$ fits.
In the $\CModelAbbrev$ model, whether to broadcast or unicast depends on a node knowing the port of the neighbour it intends to get the message to, which may be nontrivial.

The concept of compact routing was pioneered by a number of well known papers{~\cite{santoro1985labelling,PelegU88,Cowen01}}, trading stretch (factor increase in routing length) for memory used.
Several papers followed up with improvements on the schemes{~\cite{ThorupZ01,RoutingFraigniaudG01-short,Chechik13-short}}.
These schemes are made of two parts: a local preprocessing phase in which labels and routing tables are computed,
and a distributed phase in which the actual routing algorithm runs. Thus, these algorithms are in a sense not fully distributed but could be made distributed by making the preprocessing distributed.
Starting with~\cite{AwerbuchBLP90, ElkinN16a, GavoilleGHI13, LenzenP13, LenzenP15}, there has been research on efficient implementations of the preprocessing phase in the CONGEST model (e.g.~\cite{AwerbuchP92}).
However, these algorithms may not be compact (i.e. using $o(n)$ local memory)
even though the resulting routing scheme may be compact.
Elkin and Neiman~\cite{ElkinN17} claim a fully compact distributed routing algorithm using low memory in both the preprocessing and routing phases. However,~\cite{ElkinN17} and related solutions are probabilistic. Our solution, on the other hand, is deterministic
while also handling self-healing routing despite node failures, though we only handle tree based routing schemes in this work (unlike~\cite{ElkinN17} etc).

Finally, dynamic network topology and fault tolerance are core concerns of distributed computing~\cite{Attiya-WelchBook,Lynchbook} and various models (e.g.~\cite{Kuhn-DistComputation-STOC10}) and topology maintenance and  self-* algorithms abound~\cite{Djikstra74SelfStabilizing, DolevBookSelfStabilization,KormanKMPODC11,Kuhn2005Self-Repairing,KuttenPorat-DynSPanningTree-DC99,COLLIN-DolevDFS,KuttenTrehanDFS2014,Kniesburges11Re-Chord,FeldmanSST-SSDeBruijn-2017}.

\section*{Conclusions and Acknowledgements}
In this work, we formalise and give algorithms for networks of low memory machines executing streaming style algorithms developing the first fully compact self-healing routing algorithm. The power of the $\CModelAbbrev$ model we introduced needs to be studied in more detail. Algorithms in the asynchronous $\CModelAbbrev$ model and more efficient/optimal versions should be developed. We are very thankful to Danny Dolev for all the discussions and inspiration for this research.

{
\bibliography{selfheal-routing}
}

\end{document}